\documentclass[a4paper]{article}
\usepackage{graphicx}

\pdfoutput=1

\def \ins#1#2#3#4#5#6 {
  \begin{figure}[#1]
    \begin{center}
\includegraphics[width=#3,height=#4,clip]{#2}
      \caption{#5}
      \label{#6}
    \end{center}
  \end{figure}
  }

\begin{document}

\begin{center}
{\Large \bf
WEB Portal for Monte Carlo Simulations in High Energy Physics -- HEPWEB
}
\end{center}

\begin{center}
E.I. Alexandrov, V.M. Kotov, V.V. Uzhinsky, P.V. Zrelov\\
Laboratory of Information Technologies\\
Joint Institute for Nuclear Research\\
Dubna, Russia, 2011
\end{center}

\begin{center}
\begin{minipage}{14cm}
A WEB-portal HepWeb allows users to perform the most popular calculations in
high energy physics -- calculations of hadron-hadron, hadron-nucleus and nucleus-nucleus
interaction cross sections as well as calculations of secondary particles characteristics in
the interactions using Monte Carlo event generators. The List of the generators includes
Dubna version of the intra-nuclear cascade model (CASCADE), FRITIOF model, ultra-relativistic
quantum molecular dynamic model (UrQMD), HIJING model, and AMPT model.

~~~~~Setting up the colliding particles/nucleus properties (collision energy, mass numbers and
charges of nuclei, impact parameters of interactions, and number of generated events)
is realized by a WEB interface. A query is processed by a server, and results are
presented to the user as a WEB-page.

~~~~~Short descriptions of the installed generators, the WEB interface implementation
and the server operation are given.

\end{minipage}
\end{center}

\vspace{0.5cm}

\section*{Introduction}
There are some servers for on-line calculations in particle physics and nuclear physics.
The most advanced one is a NRV-server (Nuclear Reactions Video)
\cite{NRV} at the Joint Institute for Nuclear Research (JINR, Dubna, Russia). The NRV is a
system of management and graphical representation of nuclear data and
computer simulations of low energy nuclear dynamics. It consists of a complete and
renewed nuclear database and well known theoretical models of low energy nuclear
reactions altogether forming the "low energy nuclear knowledge base". The NRV
solves two main problems: fast search for and visualization of experimental data on
nuclear structure and nuclear reactions, and on-line application of commonly used
models of nuclear dynamics.
The JavaScript technology is used for forming and managing the nuclear database,
and running some nuclear models. Thus results are directly accessible through the
net with any computer having a Web-browser supported Java codes.

JetWeb-server (a WWW interface and database for Monte Carlo tuning and validation)
was created in 2002 \cite{JetWeb}. The aim of
the project was "to allow rapid and reproducible comparisons to be made between
detailed measurements at high-energy physics colliders and general physics simulation
packages". The "general" purpose Monte-Carlo simulation programs were PYTHIA \cite{Pythia}
and HERWIG \cite{Herwig}. Only HERA, LEP and Tevatron experimental data were included.

At the centre of JetWeb there was a JAVA object model containing the properties and
interactions of Models, Papers, Plots and Fits. The data underlying this model were
stored in the JetWeb database - a MySQL database. A "Model" completely specified a
unique generator, version and set of parameters. A "Paper" encapsulated the measured
data from a single publication and was associated with measured cross sections. A
"Fit" contained the results of a comparison between real data and predictions
of a specific model. The data and MC results were processed with the help of HZTOOL
package \cite{HZTOOL}, and output histograms in the form of XML files in the Java
Analysis Studio (JAS) \cite{JAS} plotML DTD format \cite{DTD} were written.
The plotML data were converted into the Java object model, and written to the JetWeb
database via JDBC \cite{JDBC}. The user accessed JetWeb via the web interface,
which consisted of Java servlets run on a Tomcat \cite{Tomcat} server, delivering
HTML pages written using the JetWeb HTMLWriter facility (ucl.hep.jetweb.html package).
The servlets accessed the JetWeb database via the JDBC calls encapsulated within the
java object model.

If a fit requested by the user were done before, the static webpages
were searched for and send to the user. If there were no results stored for the user's
specified set of parameters, a new job request with the required parameters was generated.

Now JetWeb does not operate. It seems to us that the matter is that the HZTOOL library of
the experimental data and corresponding FORTRAN routines has not been update for a long time.
As the data set was restricted, all calculations were performed quite fast,
and there was no need and possibilities to make new ones.

We think that the main idea of the JetWeb server was quite interesting, and we use them
at the creation of our server the HepWeb server.

The other interesting server is "$Q_T$ resummation portal at Michigan State University
(http://hep.pa. msu.edu/wwwlegacy/) created at 2003 \cite{QTsummed}. It allows one
to plot transverse momentum distributions for cross sections of several particle
reactions. The following processes are implemented there:
\begin{itemize}
\item Massive vector boson production --
      $pp\rightarrow W^{\pm }X$, $pp\rightarrow Z^{0}X$;
\item Photon pair production -- $pp\rightarrow \gamma \gamma X$;
\item $Z$-boson pair production -- $pp\rightarrow Z^{0}Z^{0}X$;
\item SM Higgs boson production -- $pp\rightarrow H^{0}X$.
\end{itemize}

The output figure shows distributions $d\sigma /dQ^{2}dydq_{T }$ for the
production of on-shell particles (or pairs of on-shell particles in the case of
the $\gamma \gamma$ and $ZZ$ production) with specified invariant mass $Q$,
rapidity $y$ and transverse momentum $q_{T}$ in the lab frame
(the center-of-mass frame of the hadron beams). A user can obtain resummed, fixed-order
and asymptotic cross sections. Details of program implementation is unknown for us.

In Sec. 1 below we consider  tasks of on-line calculations and formulate the aim of
our project. A description of how to use our server for Monte Carlo calculations
in CASCADE, FRITIOF, UrQMD and HIJING models is given in Sec. 2. Technical details
of the portal implementation are considered in Sec. 3. Appendices present short descriptions of the
installed generators.

\section{Task of on-line calculations}
Monte Carlo event generators play a very important role in high energy physics. One
can mark the following areas of their application:
\begin{enumerate}

\item {\bf Pragmatic or practical tasks} -- development of new or upgrade of old
experimental setups to study some processes/interactions, design of detectors,
Monte Carlo simulation of the detector responses and so on.  Event generators
applied at these should have fast operation speed, stability of work, and a rough
reproduction of previous experimental results. As an example, let us mention
applications of the UrQMD model \cite{UrQMD1,UrQMD2} for the development of detectors
for research on nucleus-nucleus interactions (CBM collaboration \cite{CMB-home-page}),
and detectors for investigation of antiproton-proton annihilations (PANDA collaboration
\cite{PANDA-home-page}) at future GSI accelerators. The RQMD \cite{RQMD} and HIJING
\cite{Hijing1,Hijing2} models have been used for analogous purposes for RHIC experiments.
The well known Geant4 package \cite{GEANT} is widely used to simulate various
installations.

\item {\bf Analysis of new experimental data and planning of new investigations }. They
include comparison of new data with previous data and model predictions. As a rule,
the new data do not agree with model predictions, so  some questions arise in this
regard: whether all special features of the setup have been taken into account;
whether they are free of methodical errors; whether the theoretical models are used
correctly; whether the model parameters set is right; whether the discrepancy between
the experimental data and model predictions is of systematical character; whether
the discrepancy was observed in previous experiments; whether the discrepancy
was considered as an evidence of a new physical effect, and so on. They are
solved differently, and often quite difficult using quite often event generators.
The generators should have a flexibility in parameters variation and physical scenario,
as well as a sufficient physical meaning of the parameters.

~~~~~Another situation takes place at new research planning. A first question asked by
experimentalists is related to the load of the setup by ordinary background processes.
A second deals with the radiation condition of the experiment. A third question asks
about the admixture of the background processes in the phenomenon under study, and how
it can be damped, and so on. Clearly, experimentalists prefer to use well approbated
and well recommended models for their  estimations. Here one can not avoid a study
of model application experience. As a rule, there is no enough time to do it. Thus,
experimental collaborations attract the authors of the models or use authors variants
of model code to solve the questions. As an example, let us point out on the estimations
of secondary particles multiplicity in central gold-gold interactions obtained by CBM
collaboration \cite{CBM_TPR}, and the estimations of background processes intensities by
the PANDA collaboration \cite{PANDA_TPR}.

\item {\bf Scientific or cognition aims only} -- search for new effects or phenomena on
the base of analysis of a discrepancy between experimental data and model predictions.
One uses the fact that Monte Carlo models are a synthesis of existing notions about
process mechanics. Thus, the discovered discrepancy can be considered as an evidence
of our insufficient understanding or else as an evidence of new effects. For example,
the discrepancy between experimental data and intra-nuclear cascade model calculations
growing with the collision energy rising led in its time to appearing a very important
conception for high energy physics -- "formation time of secondary particle".

\end{enumerate}

The final aim of all the efforts is creation of a theory of processes that could predict
effects with any predetermined exactness. As there are only few such theories, the
aim is re-formulated -- creation of a theory or a model predicting observable effects
with specified exactness. Determination of the exactness is a special additional task.

The philosophical aspect of the scientific research -- the "cognition of Good wisdom", is
out of the scope of our consideration.

The aim of this work is to create a WEB-portal which allows one to perform calculations
in the second stream.
\ins{cbth}{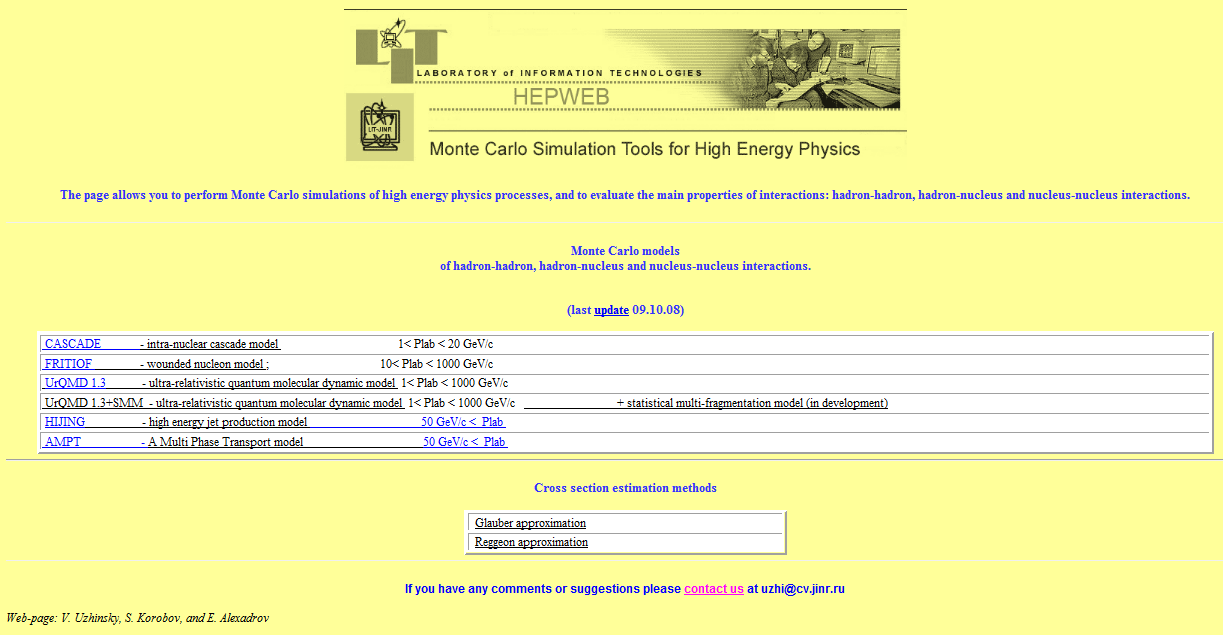}{160mm}{80mm}{View of the main page.}{mpage} 

Though the variety of tasks of the second stream is quite bright one can select out
some common tasks. We think, a user can be interested first of all in global properties
of interactions such as interaction cross sections, charged particles multiplicity
distributions, momentum, angular and energy distributions of particles. There is a
quite defined need in theoretical calculations of impact parameter distributions,
multiplicity distributions of intra-nuclear collisions, and multiplicities of wounded
nucleons. It is desirable for check of a model quality to know whether energy,
momentum, baryon number, lepton number and so on are saved in the model or not.

At high energies total, inelastic and diffraction cross sections are usually calculated
in the Glauber approach. The reggeon theory can be used also for the same aim.
All of the possibilities are offered by our server. There is a possibility in the
Glauber approximation to calculate impact parameter distributions, multiplicity
distributions of intra-nuclear collisions, and multiplicities of wounded nucleons.

The most easiest way to calculate the inclusive and global properties of interactions
is utilization of event generator programs. There are a lot of event generators
especially at low and intermediate energies which are of great practical interests.
Most of them are not accessible. It is impossible to review the left part of them. At
high energies a set of the generators is not so large. A list of most actual ones is
quite restricted. At energies below 10 GeV/nucleon various variants of intra-nuclear
cascade model are very popular. We have installed at out server the variant of CASCADE
code \cite{Zhenis} quite well known in JINR. At high energies the FRITIOF model
\cite{Fritiof} was applied quite well in the past. At present time the ultra-relativistic
quantum molecular dynamics model (UrQMD \cite{UrQMD1,UrQMD2}) is applied very often.
At super high energies, the HIJING model \cite{Hijing1,Hijing2}, PYTHIA \cite{Pythia} and HERWIG
\cite{Herwig} models are utilized. The PYTHIA and HERWIG can be applied only for
simulation of elementary particles interactions. In order to simulate nucleus-nucleus
interactions most actual now we have installed at our server the FRITIOF, UrQMD and
HIJING models. The other models will be included in future whether according to user
wishes, or mastering of new codes.

The models selected by us are quite well documented for experienced user as a rule.
Though we think that most of the potential users use the model episodically. For them
we have created a simplified access to the generators using WEB-based interface.

The main page of the portal looks like it is presented in Fig.~1. The address in INTERNET
is: http://hepweb.jinr.ru/

\section{User guide}
\subsection{Cascade Evaporation Model(CASCADE)}
The page presented in Fig.~2 gives a user a possibility to calculate properties of
interactions using well known cascade-evaporation model \cite{cascade1}--\cite{cascade6}.
A short description of the model is provided by the page also. The user should set up
mass numbers and charges of colliding nuclei, an interval for the impact parameter sampling,
a momentum of a projectile nucleus per nucleon in the target nucleus rest frame, a number of
generated events, and run the calculation process pushing "Get Results" button. All of these are
prompted by an interface presented in Fig.~2.
\ins{cbth}{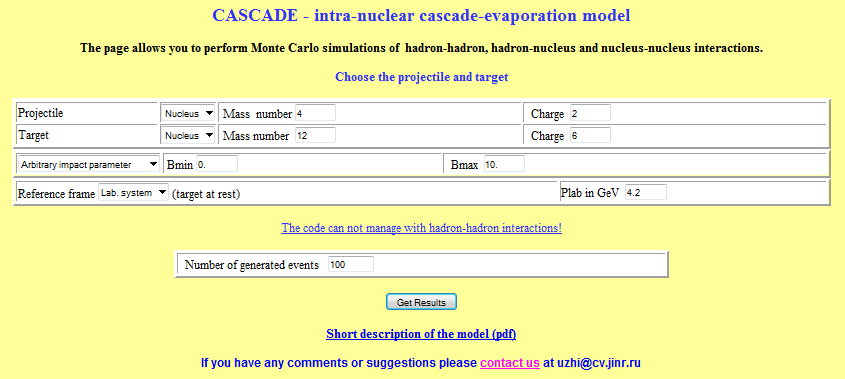}{160mm}{80mm}{View of the CASCADE page.}{cascade}

If the user is going to study hadron-nucleus interactions, he/she has to choose in the "Projectile"
list a needed item. The user should obligatory set up a mass number and a charge of the hadron.
The mass number is treated in the case as a number of particles in a projectile. Thus it
must be equal to one.

If the user is going to study minimal bias interactions, he/she can use the following
consideration for a choosing of the impact parameter interval. The radius of a nucleus with mass
number $A$ is near to the value $ R_A\sim 1.2~A^{1/3}~$ (fm). Thus for interactions
of nuclei with mass numbers $P$ and $T$ the maximum impact parameter can be estimated
as $R_P+R_T+2$ (fm). "2" represents a double radius of $NN$-interactions. The minimal
value of the impact parameter is obviously equal in the case to 0.

For a projectile hadron the user can set up $R_P$ to 1 (fm) (nucleon radius).

If you know experimental cross section of the interactions, you can estimate
the maximum value of the impact parameter as $\sqrt{\sigma_{exp}/\pi}$ (fm).

In other cases you should undertake additional consideration. A Glauber cross section
calculations can be useful for you at this (see below).

If you set up a small number of events, and are interesting by light nuclei interactions,
the results can be obtained quite fast. They will be presented on the page described
below.
\ins{cbth}{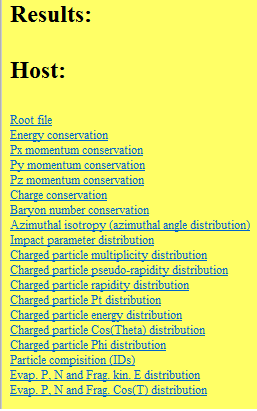}{80mm}{80mm}{View of the Results page.}{results}

If your calculations require a lot of time, the view of the page does not change
immediately. This means that your task is put in a queue for execution, or is
executing now. In the following you have two possibilities: to be at the page refreshing
it periodically until the results will be ready, or left the page and turn to your busy.

The results of your task execution will be stored in a data-base. To reach an access to them
in a new connection you should set up one more just the same parameters of the task
(mass numbers and charges of nuclei, impact parameter interval, momentum of the
projectile, number of events), and should run process as it was described before.
The server looks first of all though the data-base. If the corresponding results are
in the data-base, they will be presented on a new page\footnote{$^)$ Unlike the other
portals we do not undertake any efforts to identify users and their tasks. We are
thinking that all calculations are unique ones, and their results are important.}$^)$.

The calculation results are presented on a page, a view of which is given by Fig.~3.
In the upper part of the page, interaction parameters given by a user are presented.
After that a list of calculated interaction characteristics are followed. At clicking on
an item of the list an additional window will be opened with a graphical representation of
a characteristic. The page give you a possibility to download the results  as a
ROOT files.

The version of the cascade-evaporation model developed by Zh.Zh.~Mu\-sul\-man\-bekov
\cite{Zhenis} is used at the calculations. A short description of the model extracted
from Ref.~\cite{Adamovich} can be found at the CASCADE page.

\subsection{FRITIOF Model}
A page presented in Fig.~4 gives you a possibility to calculate properties of
interactions using the famous FRITIOF model \cite{Fritiof}. The code of the model
was essential modified by V.V.Uzhinsky \cite{UzFri}: calculations of cross sections and a sampling
of interacting nucleons within the Glauber approximation were added; an unit for
a simulation of the nuclear destruction at the fast stage of interactions (the reggeon nuclear
destruction model\cite{Khaled1}) was added also; the evaporation of nuclear residues
was taken into account and so on. Most of the changes are described in
Ref.~\cite{Adamovich}. An extraction from the paper \cite{Adamovich} related to the model is presented on
the page.
\ins{cbth}{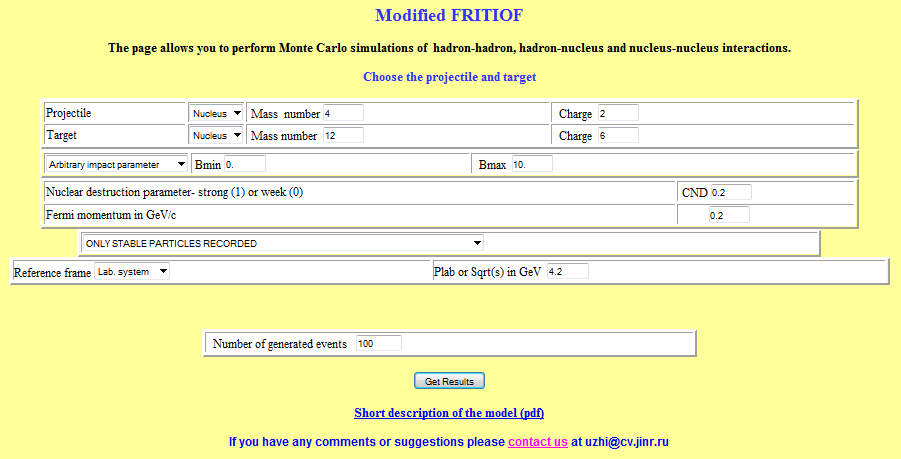}{160mm}{80mm}{View of the FRITIOF page.}{fritiof}

To run the simulation process you should set up:
\begin{itemize}
\item Mass numbers and charges of colliding nuclei, or choose a projectile particle from
      a pop-up menu (in the case arbitrary mass and charge are needed also);
\item If you are interested in the minimal bias interactions miss the next line. In other case,
      choose from a pop-up menu the item -- "Impact parameter in the range", and set up minimal
      and maximal values of the interaction impact parameter. They must be meaningful, see the
      previous section.
\item We do not recommend to change the item -- "Nuclear destruction parameter", without a special
      need. It allows to variate the nuclear destruction power at the fast stage of interactions.
      The prompted value is determined at an analysis of heavy nuclei interactions. For light
      nuclei interactions it can be changed to 1.
\item The item -- "Fermi momentum", allows you to fit the Fermi-momentum for a description, for
      example, the spectator fragment spectra.
\item The following pop-up menu gives you a possibility to order what to do with unstable
      particles at the end of the simulation of the fast stage of interactions.
\item Particle characteristics can be calculated in the laboratory system, or in the
      centre-of-mass system. Depending on your choice, set up a momentum per nucleon of the
      projectile nucleus, or the corresponding energy of the $NN$-interactions.
\item It is natural that you should set up a number of the generated events.
\item At the end of your typing, click on the button -- "Get Results".
\end{itemize}

Calculated characteristics will be presented as it was described before.

\subsection{The Ultra-relativistic Quantum Molecular Dynamics Model (UrQMD)}

The page presented in Fig.~5, gives you an opportunity to calculate
the interaction characteristics using UrQMD model \cite{UrQMD1,UrQMD2}. Changes, made in the
original code of the model, are described in details in Ref.~\cite{UzhiUrQMD}.
\ins{cbth}{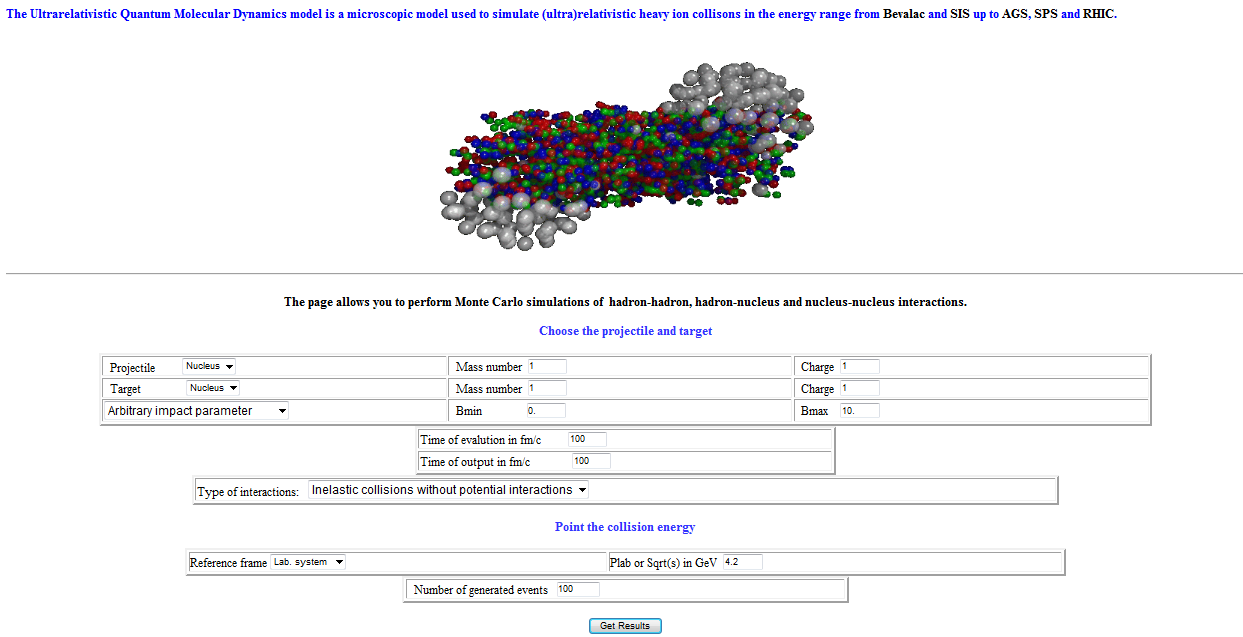}{160mm}{100mm}{View of the UrQMD page.}{urqmd}

To run the simulation process you should set up:
\begin{itemize}
\item Mass numbers and charges of colliding nuclei, or choose a projectile particle from
      a pop-up menu (in the case arbitrary mass and charge are needed also);
\item If you are interested in the minimal bias interactions miss the next line. In other case,
      choose from a pop-up menu the item -- "Impact parameter in the range", and set up minimal
      and maximal values of the interaction impact parameter. They must be meaningful, see the
      previous section.

\item We do not recommend to change the items -- "Time of evolution in fm/c", and
      "Time of output in fm/c" without defined need. They determine a time of interactions
      and a frequency of an intermediate results output. It is obvious that the time of evolution
      must not be lower than the time of a particle penetration through a target nucleus. At low and
      intermediate energies the change of the parameter has an influence on properties of
      the pre-equilibrium particles and residual nuclei. Giving a small time of the evolution,
      you will obtain a "hot" residual nucleus with a large mass because the pre-equilibrium
      particles will be captured by the residual. At large evolution time the nucleus will
      be more "cold", and the multiplicity of the pre-equilibrium particles will be larger.

\item The item -- "Type of interactions", allows you to order to perform calculations with or
      without potential interactions. The corresponding sub-menu items are "Inelastic collisions
       + potential interactions" and  "Inelastic collisions without potential interactions".
       At energies larger then 4 GeV/nucleon all calculations are performed without the potential
       interactions.

\item Particle characteristics can be calculated in the laboratory system, or in the
      centre-of-mass system. Depending on your choice, set up a momentum per nucleon of the
      projectile nucleus, or the corresponding energy of the $NN$-interactions.

\item It is natural that you should set up a number of the generated events.

\item At the end of your typing, click on the button -- "Get Results".

\end{itemize}

Calculated characteristics will be presented as it was described before.

Below the menu there is a reference on the UrQMD model validation WEB-page
which collects materials on a comparison of the UrQMD model calculations and
experimental data \cite{Uzhi_Val}.

\subsection{HIJING Model}

The page presented in Fig.~6, gives you an opportunity to calculate
properties of hadron-hadron, hadron-nucleus, and nucleus-nucleus interactions at high
and super high energies using the HIJING model \cite{Hijing2}. We use a variant of the code
disposed at the code depository of the LCG project \cite{GENSER} and described in
Ref.~\cite{Uzhi_HIJVal}.
\ins{cbth}{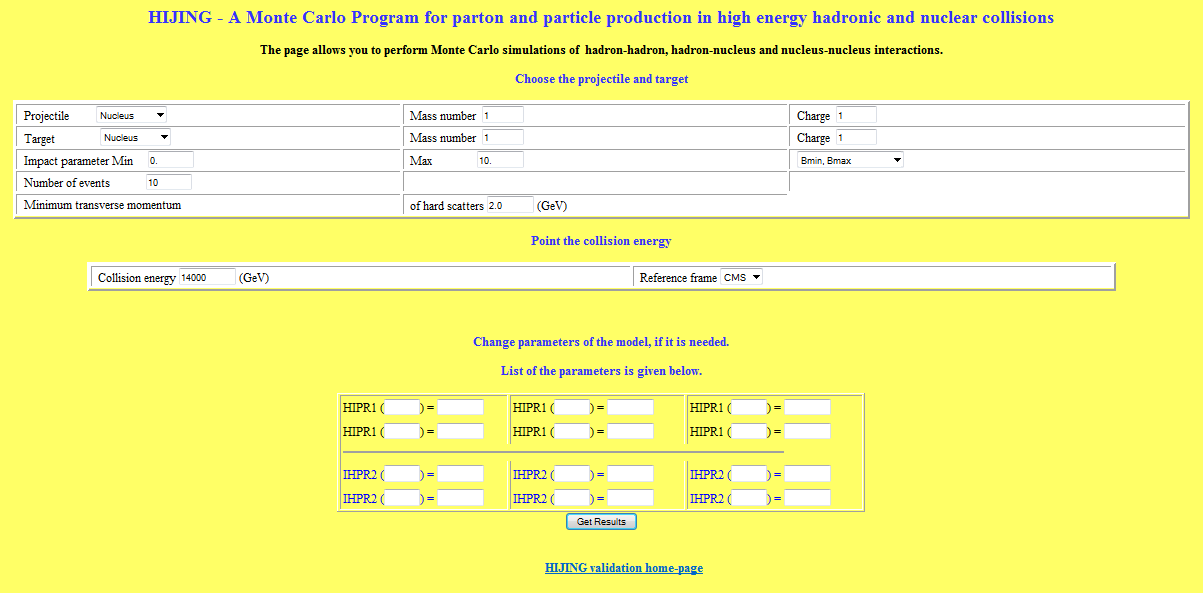}{160mm}{80mm}{View of the HIJING page.}{hijing}

To run the simulation process you should set up:
\begin{itemize}
\item Mass numbers and charges of colliding nuclei, or choose a projectile particle from
      a pop-up menu (in the case arbitrary mass and charge are needed also);

\item If you are interested in the minimal bias interactions miss the next line. In other case,
      choose from a pop-up menu the item -- "Impact parameter in the range", and set up minimal
      and maximal values of the interaction impact parameter. They must be meaningful, see the
      previous section.

\item Set up a number of generated events.

\item We do not recommend to change one of the most important parameters of the model --
      "Minimum transverse momentum of hard scatters", if you have not a special need.

\item Particle characteristics can be calculated in the laboratory system, or in the
      centre-of-mass system. Depending on your choice, set up a momentum per nucleon of the
      projectile nucleus, or the corresponding energy of the $NN$-interactions.

\item We give in the menu an additional possibility to change the model parameters and options.
      A list of the parameters and options is given below. For example, to switch off the jet
      quenching effect it is need to put IHPR2(5)=0. This means that you should enter the
      corresponding windows numbers "5" and "0".

\item At the end of your typing, click on the button -- "Get Results".

\end{itemize}

Calculated characteristics will be presented as it was described earlier.

Below the button -- "Get Results", there is a reference on the HIJING model validation
WEB-page \cite{Uzhi_HIJVal} which collects materials on a comparison of the HIJING
model calculations and experimental data.

\subsection{Glauber cross section calculation}

A page presented in Fig.~7 gives you an opportunity to calculate hadron-nucleus and
nucleus-nucleus interaction characteristics in the Glauber approximation such as
the total cross section, the elastic cross section, the AB$->$XB cross section, the AB$->$AX cross section,
the AB$->$X cross section, the production cross section, the production cross section of interactions
with composite materials, and various distributions of inelastic reactions. For this you
should choose needed item and click a button -- "Run".
\ins{cbth}{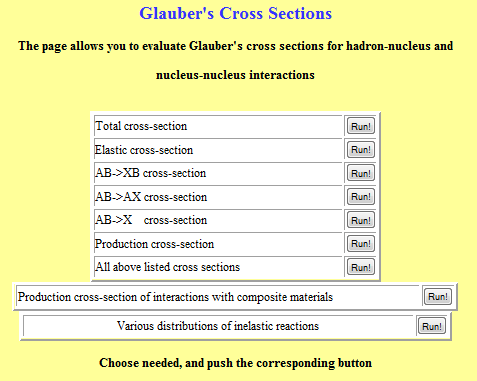}{160mm}{110mm}{View of the Glauber page.}{glauber}

All following pages for the Glauber calculations have nearly the same view presented in Fig.~8.
To perform a calculations you should set up:
\begin{itemize}
\item Mass numbers and charges of colliding nuclei. If you are going to study
      hadron-nucleus interactions, set up for the mass and charge of the projectile
      the values "1" and "1".

\item After that you should set up $NN$-interaction properties -- the total interaction cross
      section, the slope of differential elastic scattering cross section, and the ratio of real
      and imaginary parts of the elastic scattering amplitude at zero momentum transfer.
      All of the quantities can be found in the compilation \cite{compil}. The same
      quantities can be calculated assuming the gaussian parameterization of the elastic
      scattering amplitude and $Re F/Im F=0$ if you know the total and elastic $NN$
      interaction cross sections. For a determination of the $NN$ cross sections one can
      use parameterizations proposed by the Particle Data Group (PDG \cite{PDG}). We
      allow both of the possibilities. To choose one of the methods you should mark
      one of them in the corresponding check box.

      If you choose an usage of the PDG paramerization, you should additional set up
      a reference frame and a momentum per nucleon or an energy in the centre-of-mass system
      ($\sqrt{s}$).

      You need to set up the required parameters by yourself if you are going to consider
      hadron-nucleus interactions ($\pi A,~ KA$).

\item In the codes the stochastic method of an evaluation of the multi-dimensional integrals is used.
      It is quite well described in Refs. \cite{Zador, DIAGEN, Galperin}. Thus, you have
      to set a number of samples (statistic). It determines an accuracy of the calculation.

\item At the end of your typing, click on the button -- "Get Results".

\end{itemize}
\ins{cbth}{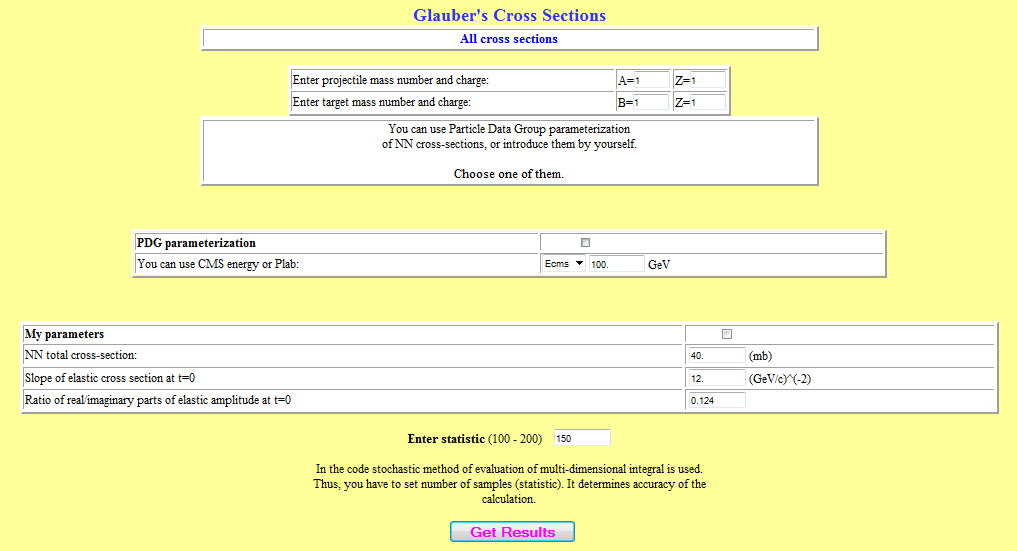}{160mm}{110mm}{View of the Glauber calculation page.}{allX}

\subsection{Reggeon cross section calculation}

The page presented in Fig.~9 gives you an possibility to calculate hadron-hadron,
hadron-nucleus, and nucleus-nucleus interaction properties using the reggeon or pomeron
approach.
\ins{cbth}{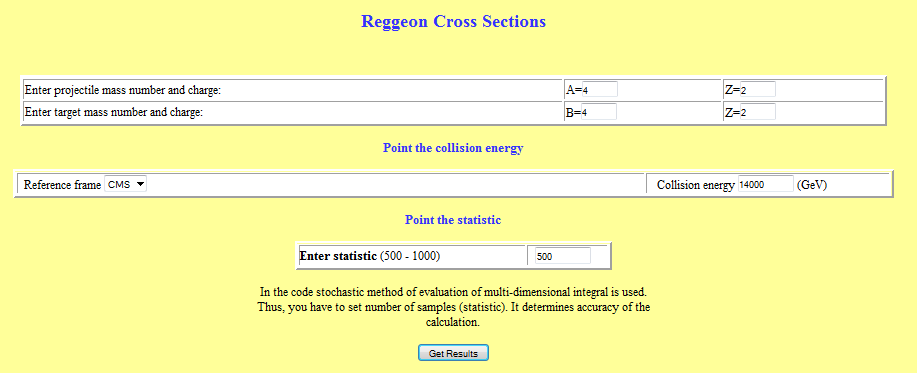}{160mm}{80mm}{View of the Reggeon page.}{reggeon} 

To perform a calculations you should set up:
\begin{itemize}

\item Mass numbers and charges of colliding nuclei. If you are going to study
      hadron-nucleus interactions, set up for the mass and charge of the projectile
      the values "1" and "1". If you are interesting in nucleon-nucleon collisions
      set up also for the target the values "1" and "1" (Proton is a nucleus of hydrogen
      atom.)

\item You should set up a reference frame and a momentum per nucleon, or an energy of
      $NN$-interactions in the centre-of-mass system. They are needed for a calculation
      of the reggeon parameters.

\item Because in the code the stochastic method of an evaluation of the multi-dimensional
      integrals is used, you have to set a number of samples (statistic). It determines
      an accuracy of the calculation.

\item At the end of your typing, click on the button -- "Get Results".
\end{itemize}

A description of the calculating method will be published elsewhere.

\section{Structure of WEB-service HepWeb}
Already existing servers described in the Introduction have been analyzed,
and the following requirements to the HepWeb service have been formulated:
\begin{enumerate}
\item The HepWeb service should have a WEB-oriented architecture;
\item Requirements to a user's computer should be minimal one both on hardware and software
      levels;
\item Generators should have standardized input and output streams;
\item Generators satisfying to the point 3 should be easily installed and
      modified on the server, i.e. they should not have any special
      settings necessary only for the HepWeb;
\item Result of a generator work should be presented in standard graphic
      formats with possibility to create a root file or text fragments;
\item Result of the generator work should be saved because some calculations
      require a lot of a computational time, and it is inexpedient to repeat them.
\end{enumerate}

The following structure was accepted and implemented in the accordance with the
given above requirements:
\begin{itemize}
\item The HepWeb service is based on the "client - server" architecture.
\item The HepWEb service is implemented in Java, and uses the JavaServer Pages
      (JSP) technology. This allows one to result as XHTML - or
      XML-documents. User can apply any Web-browser (IE, Mozilla, Opera etc.)
      for their viewing.
\item The output streams of all HepWeb generators are standardized.

An output data is written in a text file. Any number of histograms can be stored in
the file. A histogram has a header where a name of the histogram and the names
of the co-ordinate axes are given. Then a set of co-ordinates is followed.
Additionally a ROOT-file of the histograms could be also created.

\item The results are stored in the database.
\end{itemize}

All generators are presented on the server as executable modules (the source files of
the codes are unavailable for users). The modules can be also started independently
of the HEPWEB. This allows one to upgrade generators without reboot of the server
and add new generators specifying only ways to the corresponding executable modules in
a configuration file.

The standardized output stream is analyzed by a special method created by us
in order to extract the histograms. The histograms are converted into graphic files
(png-files) using the standard Java-technology.

The MySQL database was chosen to store the results because it is free, and its
bandwidth of the read/write data is enough for the HepWeb service.

A interaction scheme between the components of the HepWeb service in presented in Fig. 10.
\ins{cbth}{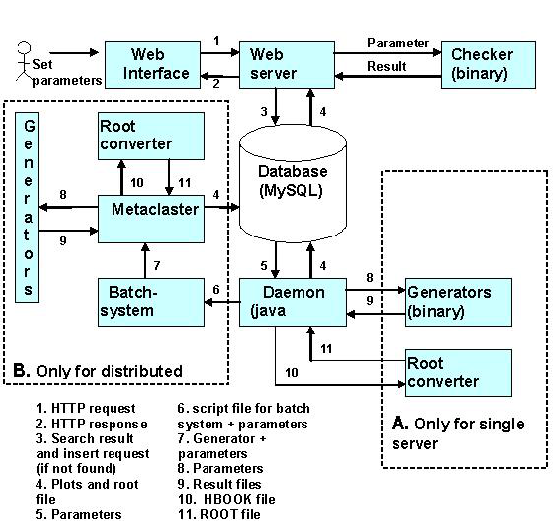}{160mm}{100mm}{Structure of the HepWeb-server.}{Zhenya} 

A client generates a request to the server using a form which is built in
a static HTML page. Each generator uses its own HTML page with an according form. An answer
on the query is generated by the server with a help of the JSP technology. The server
checks correctness of the parameters of the query. For example, it is checked that
the energy of $NN$-collision in the centre-of-mass system $ \sqrt {s} $ is larger
than the sum of nucleon masses, or that the range of the impact parameter does not extend
to negative values, etc. All correct requests are stored in the database. The server
checks for each request if a request is already stored in database or not. If the
request is stored, then the server generates the result page and  sends it to the user.
If the results for the request are not presented, an empty page is sent to the client
with a line -- "Please wait a few minutes then refresh this page".
In the case the requested generator with corresponding parameters is started on the server.

The HepWeb service has two possibilities for executing the generator:
\begin{enumerate}
\item  Use only one computer for the server;
\item  Use distributed computing system for the server.
\end{enumerate}

\subsection{The single-computer based HepWeb service structure}
For the structure, all calculations are performed by one computer. A request
to run a generator follows to the consecutive list. Thus, all queries of a generator
will be executed in consecutive order. The given structure is efficient for a small number of
the requests. If the number of queries increases, the astronomical time of their
execution will increase too. The structure is marked by letter "A" on Fig. 10.

\subsection{Structure of the HepWeb service based on a distributed computer system}
The structure is developed for starting tasks on several computers. It allows one to
perform several queries simultaneously. In the presence of a sufficient quantity
of computers in the system, the performance time of a new query is nearly equal to
the performance time of this task on one computer. This structure is marked by letter
"B" on Fig. 10.

The HepWeb service was tested in 2006 on distributed metaclaster of the Dubna-Grid project (\cite{LITSR2005_1},\cite{LITSR2005_2}).
An interaction of the service in batch-mode of access to resources of the metaclaster
has been successfully proved.

\section*{Summary}
A WEB-portal is designed which allows a user to perform the most popular calculations in
high energy physics -- calculations of hadron-hadron, hadron-nucleus and nucleus-nucleus
interaction cross sections and calculations of secondary particles characteristics in
the interactions using Monte Carlo event generators. The list of the generators includes
the intra-nuclear cascade model (CASCADE), the FRITIOF model, the ultra-relativistic
quantum molecular dynamic model (UrQMD), and HIJING model.

Setting up of colliding particles properties (collision energy, mass numbers and
charges of nuclei, impact parameters of interactions, and a number of generated events)
is realized by a WEB interface. A query is processed by server, and results are
presented to the user as a WEB-page.

\section*{Appendix: The Cascade-Evaporation Model (CEM)}
The cascade-evaporation model (see Refs. \cite{cascade1}--\cite{cascade6}) was one of the first
models of hadron-nucleus interactions at high energies ($E>1$ GeV) very important for practical
applications. It saves its position at present time too. The model assumes that due to an inelastic interaction
of a projectile hadron with one of target nucleons a new particle is produced.
The participating target nucleon accepts momentum and begins moving in the nucleus. All moving
(cascade) particles can interact with other nuclear nucleons to produce new particles or suffer
an elastic re-scattering. Therefore, a cascade reproduction of moving particles is assumed. The
interactions between the cascade particles are usually omitted. The process continues until all
moving particles either leave the nucleus or are absorbed.

In the case of nucleus-nucleus collisions, it is assumed that the cascade particles can interact
with projectile and target nucleons. To choose the nucleons which can participate in elementary
interactions, all of the nucleons of the colliding nuclei with mass numbers A and B are identified
by the coordinates ($x_{i}, y_{i}, z_{i},\ 1 \le i \le  A)$  and  $(x_{j}',y_{j}', z_{j}',\  1 \le  j \le  B)$
in the initial state in the corresponding reference frames. At
the sampling of the nucleon coordinates for a nucleus with A (B) $\le  10$ the oscillator density
is used. The Wood-Saxon density with the parameters $R_{A}=1.07*A^{1/3}$ fm and $c=0.545$ fm
is applied for more heavy nuclei. The nucleon core is taken into account at sampling
the nucleon coordinates according the the densities $\rho _{A}$ and $\rho_{B}$ (two nucleons
can not be be closer than $R_{c}$, $R_{c}=0.4$ fm).

Taking into account the Lorentz contraction of the projectile (A) in the rest frame of the target
nucleus, the corresponding coordinates are redefined as $z_{i} \rightarrow z_{i}/\gamma - R_{A}/\gamma - R_{B}$.
Here $\gamma $ is the Lorentz factor of the projectile nucleus, $R_A$ and $R_B$) are radii of the nuclei.
If the coordinates of the i-th nucleon of the nucleus A and the j-th nucleon of the nucleus B
satisfy the condition
$$
(b_{x}+x_{i}-x_{j}')^{2}+(b_{y}+y_{i}-y_{j}')^{2} \le  (R_{int} +
\lambda _{D})^{2},
$$
then the two nucleons are considered as possible participants in the elementary interactions.
Here ($b_x$, $b_y$) are components of the impact parameter, $R_{int}$ is the strong interaction
radius (1.3 fm) and $\lambda_{D}$ is the de Broglie wavelength of the projectile nucleon. The
possible participant can fly without interaction, or suffer an elastic scattering, or an inelastic
interaction. The corresponding probabilities are
$$W_{tot}(\pi*(R_{int}+\lambda _{D})^{2}- \sigma ^{tot})/(\pi*(R_{int}+\lambda _{D})^{2}),$$
$$W_{el}=\sigma ^{el}/(\pi*(R_{int}+\lambda _{D})^{2}),$$
$$W_{in}= \sigma ^{in}/(\pi*(R_{int}+\lambda _{D})^{2}),$$
where $\sigma ^{tot}$, $\sigma ^{el}$  and $\sigma ^{in}$
are the total, elastic and inelastic nucleon-nucleon (NN) interaction
cross-sections. The time of possible interactions in the frame of the target nucleus is
determined as $t_{ij} = (z_j' - z_i)/v$ where $v$ is the velocity of the projectile nucleus.

A minimal time interaction will be caused first. The nucleons taking part in the
interaction are ascribed a Fermi momentum in the corresponding reference frame and the
momentum of the projectile nucleon is transferred to the target nucleus rest frame. It is
assumed that the Fermi momentum distribution has the form
\begin{equation}
W(p)dp= 3 p^{2}/P_{F}^{3}(r) dp,~~~0\le  p \le P_{F}(r), \label{FMD}
\end{equation}
where $P_{F}(r)$ depends on the local density of the nucleus $\rho (r)$,
\begin{equation}
P_{F}=\hbar(3\pi^{2}\rho(r))^{1/3}. \label{FM}
\end{equation}
$\vec r$ is the radius-vector of the interacting nucleon in the corresponding nucleus.
In the following the Fermi momenta are ascribed only to new involved nucleons.

An elementary interaction (elastic scattering or inelastic collisions) will be rejected,
if at least one of the two colliding nucleons, after the interaction, falls into the region of
occupied states of nucleus A or B (has an energy lower than the Fermi energy of $P^2_F /2m$).
In this case the next possible interaction in the time is considered. This repeats until an
interaction occurs.

The simulation procedure of elastic and inelastic nucleon-nucleon and meson-nucleon
interactions is described in detail in  \cite{cascade1}. It allows one to reproduce
the experimental observations up to energies $\sim  10$ -- 15 GeV, and especially
the particle momentum distributions.

After the first real interaction, the time is increased by $t_{ij}$. The coordinates of the
moving particles are changed as $z_k \rightarrow z_k+vt_{ij}$. The coordinates of newly produced particles
($\pi$-mesons) in the interaction with equal probabilities are identified by the coordinates of
the target or projectile nucleons ($\vec r_i$ or $\vec r_j'$). Then all possible interactions of the produced
and moving particles with the projectile or target nucleons are considered. Each cascade
particle can interact with nucleons in a tube with radius $R_{int} + \lambda _{D}$ along its trajectory.
The time for possible interactions is determined as
$t_{ik}=(\vec v_{k} (\vec r_{i}-\vec r_{k} ''))/\mid \vec v_{k}\mid$ or
$t_{jk}=(\vec v_{k} (\vec r_{j} '- \vec r_{k} '')/\mid \vec v_{k}\mid$,
where $v_k$ is the velocity of the cascade particle $k$, and $\vec r_k''$ is its radius vector.

Among all possible interactions the one with the minimum time is chosen. The
processes continue until all possible interactions have been considered.

As mentioned above, the nucleon taking part in an interaction is considered as a
cascade particle. Due to of this, the local density of the nucleus gets lower. This is
the so-called trailing effect.

At the end of the fast cascade stage of the process, the number and charges of spectator
nucleons as well as the charges of the absorbed mesons determine the nuclear residual mass
number and charge.

The excitation energy of the nuclear residual is the sum of the energies of absorbed
particles and the holes counted from the Fermi energy. It is assumed that due to an
interaction of a cascade particle or a projectile nucleon with a target nucleon having the
energy $p^2/2m_N$, where p is determined according to the distribution (\ref{FMD}), a hole with
an energy $E_h = P^2_F /2mN - p^2/2m_N$ is created in the target nucleus. ( $P_F$ is given by
equation (\ref{FM})). If an energy of the target nucleon after the interaction - $T$, is larger than
the Fermi energy ($P^2_F /2m_N$) but lower than $P^2_F /2m_N + E_b$ ($E_b$ = 7 MeV), it is assumed
that the target nucleon will not leave the nucleus. It is considered to be an absorbed
nucleon. In this case its yield to the excitation energy of the nucleus is determined as
$E_n = T - P^2_F/2m_N$. A meson having an energy lower than 25 MeV counted from the
bottom of the potential well is considered as to be an absorbed one, too. Its yield to
the excitation energy of the nucleus is determined as $T_{\pi} + m_{\pi}$. (It is assumed that the
elementary interaction is caused in the potential well with a depth $P^2_F/2m_N + E_b$. This
value is added to the kinetic energy of the incoming cascade particle or the projectile
nucleon and is subtracted from the kinetic energy of each outgoing particle.)

The total excitation energy of the target nucleus is a sum of the energies of the holes,
the energies of absorbed nucleons and mesons. The analogous procedure can be applied to
determine the excitation energy of the projectile nucleus. It is obvious that the excitation
energy of the target nucleus will be proportional to the number of ejected nucleons if one
neglects the absorption of the nucleons and mesons.

Let us note that according to the above prescription in the limit case when all
nucleons of the target nucleus are ejected one can obtain a nucleus without any nucleons,
but with a defined excitation energy.

The excitation energy governs the nuclear residual relaxation. So the method of the
excitation energy calculation links the fast and slow stages of the interaction.

The nuclear residual relaxes before thermodynamic equilibrium and can emit the so-called
pre-equilibrium particles. This process takes place if the number of quasi-particles --
$N_{q}=N_{h}+N_{n}$, is larger than the equilibrium value
$N_{q\,(eq.)}=2\sqrt{6aE^{*}/\pi ^{2}}$. Here $N_h$ is the number
of holes (the number of participating nucleons), $N_n$ is the number of absorbed cascade
nucleons, $E^*$ is the excitation energy of the nucleus, and {\it a} is the level density parameter,
taken as {\it à}$=A/10$~MeV$^{-1}$. The pre-equilibrium decay of the nucleus is simulated in the framework
of the exciton model \cite{Blann66}. The decay of a thermalized nucleus is described by the usual
evaporation approach \cite{Weis37, Fried83}.

Additional details of the model can be found in \cite{cascade6}. We have installed
at the HepWeb server a code described in \cite{Zhenis}.

From the view point of modern approaches, the realization of CEM simulation is regarded as being
too simple. It fails to take into account many important effects: a variation of the average nuclear
field during the collision, a production of meson and baryon resonances, a finite formation time of
a particle, a coalescence of nucleons, a multi-fragmentation of nuclei and so on. Thought the analogous
implementation of the model \cite{Mashnik94} was recognized as the best model code used in
the physics of the intermediate energies.

\section*{Appendix: The UrQMD Model}
A variation of the average nuclear mean field and its influence on the inclusive distributions at low
and intermediate energies are commonly considered in the VUU/BUU approaches
(Vlasov-Weling-Ulenbek or Bolzman-Weling-Ulenbek, see Refs. \cite{Bertch, Cassing}).
At higher energies or at a strong breakup of nuclei the effects of the average field render weak.
Here, the quantum molecular dynamical approach considering the explicit form of the two- and
three particle interactions is preferably applied. Its relativistic generalization combined
with quark ideas of the multi-particle production is presented in the RQMD model
\cite{Sorge1, Sorge2, Sorge3} and UrQMD model \cite{UrQMD1, UrQMD2}).  It allows one to
describe numerous characteristics of produced particles in nucleus-nucleus (AA) collisions at
high energies. The UrQMD model is well described in \cite{UrQMD1, UrQMD2}. Thus below we give
only the main features of the model.

The UrQMD model describes hadronic interactions at low and intermediate
energies ($\sqrt s <5$~GeV, $P_{lab}\leq 12$~GeV/c) in terms of interactions
between known hadrons and their resonances. At higher energies, $\sqrt s >5$~GeV,
the excitation of color strings and their subsequent fragmentation
into hadrons are taking into account.

The model is based on the covariant propagation of all hadrons
considered on the (quasi-)particle level on classical trajectories in
combination with stochastic binary scatterings, color string
formation and resonance decay. It represents a Monte Carlo solution of a
large set of coupled partial integro-differential equations for the time
evolution of the various phase space densities of particle species
$i=N,\Delta,\Lambda,$ etc. The main components of the model
are cross sections of binary reactions, potentials and decay widths of
resonances.

The potential interaction is based on a non-relativistic density-dependent
Skyrme-type equation of state with additional Yukawa- and Coulomb potentials.
Momentum dependent potentials are not used. The Skyrme potential consists of a
sum of two- and a three-body interaction terms. The two-body term, which
has a linear density-dependence models the long range attractive
component of the nucleon-nucleon interaction, whereas the three-body
term with its quadratic density-dependence is responsible for the
short range repulsive part of the interaction. The parameters of the
components are connected with the nuclear equation of state. Only the
hard equation of state has been implemented into the current UrQMD model.

The impact parameter of a collision is sampled according to the quadratic
measure ($dW\sim~bdb$). At the given impact parameter, the centers of projectile
and target are placed along the collision axis in such a manner that
the distance between surfaces of the projectile and the target is equal
to 3 fm. Momenta of nucleons are transformed in the system where
the projectile and target have equal velocities directed in different
directions of the axis. After that the time propagation starts.
During the calculation each particle is checked at the beginning of each
time step whether it will collide within that time step. A collision
between two hadrons will occur if $d<\sqrt{\sigma^{tot}/\pi}$, where $d$
and $\sigma^{tot}$ are the impact parameter of the hadrons and the total
cross section of the two hadrons, respectively. After each binary
collision or decay the outgoing particles are checked for further
collisions within the respective time step.

In the UrQMD model the total cross section $\sigma^{tot}$ depends
on the isospins of colliding particles, their flavor and the c.m. energy.
The total and elastic proton-proton and proton-neutron cross sections
are well known \cite{PDG92}. Since their functional dependence on
$\sqrt{s}$ shows a complicated shape at low energies, UrQMD uses a
table-lookup for those cross sections. The neutron-neutron cross section
is treated as equal to the proton-proton cross section (isospin-symmetry).
In the high energy limit ($\sqrt{s} \ge 5$~GeV) the CERN/HERA
parameterization for the proton-proton cross section is used
\cite{PDG92}.

Baryon resonances are produced in two different ways, namely
\begin{itemize}
\item[\bf i)] {\it hard production\/}:
N+N$\rightarrow \Delta$N,\ $\Delta\Delta$,\ N$^*$N, etc.
\item[\bf ii)] {\it soft production\/}: $\pi^-+$p$\rightarrow
\Delta^0$,\ K$^-$+p$\rightarrow \Lambda^*$...
\end{itemize}
The formation of $s$-channel resonances is fitted to measured data.
Partial cross-sections are used to calculate the relative weights for
the different channels.

There are six channels of the excitation of
non-strange resonances in the UrQMD model, namely $NN \rightarrow N
\Delta_{1232}, N N^{\ast}, N \Delta^{\ast}, \Delta_{1232}
\Delta_{1232}, \Delta_{1232} N^{\ast}$, and $\Delta_{1232}
\Delta^{\ast}$.
The $\Delta_{1232}$ is explicitly listed,
whereas higher excitations of the $\Delta$ resonance
are denoted as $\Delta^{\ast}$.
For each of these 6 channels specific assumptions were made
with respect to the form of the matrix element, and the free
parameters were adjusted to the available experimental data.

Meson-baryon (MB) cross sections are dominated by the formation of $s$-channel
resonances, i.e. the formation of a transient state of mass
$m=\sqrt{s_{hh}}$, containing the total c.m. energy of the two incoming
hadrons. On the quark level such a process implies that a quark from
the baryon annihilates an antiquark from the incoming meson. Below 2.2 GeV
c.m. energy intermediate resonance states get excited. At higher energies
the quark-antiquark annihilation processes become less important. There,
$t$-channel excitations of the hadrons dominate, where the exchange of
mesons and Pomeron exchange determines the total cross- section of the
MB interaction \cite{donna}.

To describe the total meson-meson reaction cross sections,
the additive quark model and the principle of detailed
balance, which assumes the reversibility of the particle interactions
are used.

Resonance formation cross sections from the
measured decay properties of the possible resonances up to
c.m. energies of 2.25~GeV for baryon resonance and 1.7~GeV
in the case of MM and MB reactions have been calculated based on the
principle. Above these energies collisions are modeled by the formation
of $s$-channel string or, at higher energies (beginning at $\sqrt s =3$~GeV),
by one/two $t$-channel strings. In the strangeness channel
elastic collisions are possible for those meson-baryon combinations
which are not able to form a resonance, while the creation of
$t$-channel strings is always possible at sufficiently large
energies. At high collision energies both cross sections become equal
due to quark counting rules.

A parameterization proposed by Koch and Dover \cite{kochP89a}
is used in UrQMD model for baryon-antibaryon annihilation cross section.
It is assumed that the antiproton-neutron annihilation cross section is
identically to the antiproton-proton annihilation cross section.

The hadron-hadron interactions at high energies are simulated in 3 stages.
According to the cross sections the type of interaction is defined:
elastic, inelastic, antibaryon-baryon annihilation etc. In the case of
inelastic collision with string excitation the kinematical characteristics
of strings are determined. The strings between quark and diquark
(antiquark) from the same hadron are produced. The strings have the continuous
mass distribution $f(M) \propto 1/M$ with the masses $M$, limited by
the total collision energy $\sqrt{s}$: $M_1+M_2 \le \sqrt{s}$.
The rest of the $\sqrt{s}$ is equally distributed
between the longitudinal momenta of two produced strings.

The second stage of h-h interactions is connected with
string fragmentation. The string break-up is treated
iteratively: String $\rightarrow$ hadron + smaller string.
A quark-antiquark (or a diquark-antidiquark) pair is created and
placed between leading constituent quark-antiquark (or diquark-quark)
pair.  Then a hadron is formed randomly on one of the end-points of the
string. The quark content of the hadron determines its species
and charge. In case of resonances the mass is determined according to
the Breit-Wigner distribution. Finally, the energy-fraction of the string
which is assigned to the newly created hadron is determined:
After the hadron has been stochastically assigned a transverse momentum,
the fraction of longitudinal momentum transferred from the string
to the hadron is determined by the fragmentation function.
The conservation laws are fulfilled. The diquark is permitted to
convert into mesons via the breaking of the diquark link.

This iterative fragmentation process is repeated until the remaining
energy of the string gets too small for a further fragmentation.

The fragmentation function $f(x,m_t)$ represents the probability distribution
for a hadron with the transverse mass $m_t$ to acquire the longitudinal
momentum fraction $x$ from the fragmenting string. One of the most
common fragmentation functions is the one used in the LUND model
\cite{andersson83a}.
In UrQMD, different fragmentation functions are used for leading nucleons
and newly produced particles, respectively:
\begin{eqnarray}
\label{fuqmd1}
f(x)_{\rm nuc} &=& \exp\left(-\frac{(x-B)^2}{2\,A^2}\right)
                        \quad \mbox{for leading nucleons} \\
\label{fuqmd2}
f(x)_{\rm prod} &=& (1-x)^2
                        \quad \mbox{for produced particles}
\end{eqnarray}
with $A=0.275$ and $B=0.42$.
The fragmentation function $f(x)_{\rm prod}$, used for newly produced
particles, is the well-known Field-Feynman
fragmentation function \cite{field77a,field78a}.

The fragmentation scheme determines the formation time of created hadrons. Though
there are various possibilities (for details see Refs.
\cite{UrQMD1,UrQMD2}).

After the fragmentation, decay of the resonances proceeds according to the
branching ratios compiled by the Particle Data Group \cite{PDG92,PDG}.
The resonance decay products have isotropical distributions in the rest
frame of the resonance. If a resonance is among the outgoing particles,
its mass must first be determined  according to the Breit-Wigner
mass-distribution. If the resonance decays into $N > 2$ particles,
then the corresponding $N-$body phase space is used to calculate their $N$
momenta stochastically.

The Pauli principle is applied to hadronic collisions or decays by
blocking the final state if the outgoing phase space is occupied.

The UrQMD collision term contains 55 different baryon species
(including nucleon, delta and hyperon resonances with masses up
to 2.25 GeV) and 32 different meson species (including strange
meson resonances), which are supplemented by their corresponding
anti-particle and all isospin-projected states. The states can either
be produced in string decays, s-channel collisions or resonance decays.
For excitations with higher masses than 2 GeV a string picture is
used. Full baryon/antibaryon symmetry includes:
The number of the implemented baryons therefore defines the number
of antibaryons in the model and the antibaryon-antibaryon interaction
is defined via the baryon-baryon interaction cross sections.

The elementary cross sections are fitted to available proton-proton or
pion-proton data. Isospin symmetry is used when possible in order
to reduce the number of individual cross sections which have to
be parameterized or tabulated.

We have installed a bug fixed version of UrQMD 1.3
\cite{UzhiUrQMD} at the HepWeb server.

\section*{Appendix: The FRITIOF Model}
A role of heavy resonances in hadron-hadron interactions becomes more and more essential
with a growth of a collision energy. The resonances have wide mass distributions. The
distributions overlap, and the mass spectra in the binary reactions become
practically continuous and smooth. This is taken into account in the Fritiof model \cite{Fritiof}.
The model assumes that all hadron-hadron interactions at high energies are binary reactions:
$a + b \longrightarrow a'+ b'$, where $a'$ and $b'$ are excited states of initial hadrons $a$ and $b$.
The excited states are characterized by masses which are selected by the following procedure:
in the center of mass of colliding hadrons the energy-momentum conservation low has form:
\begin{eqnarray}
E_a + E_b & = & E_{a'} + E_{b'} = \sqrt{s_{ab}}, \nonumber \\
p_{a z} + p_{b z} & = & p_{a' z} + p_{b' z} =0, \label{1}\\
0 & = & \vec p _{a' \perp} + \vec p _{b' \perp}, \nonumber
\end{eqnarray}
where $E_a$ and $E_b\ (E_{a'},\ E_{b'}) $ -- energies of initial (final) hadrons $a$ and $b\ (a',b')$,
$p_{az}$ and $P_{bz}$ -- longitudinal momentum components, $\vec p _{a' \perp}$ and
$\vec p _{b' \perp}$ -- transverse momentum components of hadrons $a'$ and $b'$.

Adding and subtracting the first two equations from (\ref{1}) leads to:
\begin{eqnarray}
P^+ _a + P^+ _b &=& P^+ _{a'} + P^+ _{b'} \nonumber \\
P^- _a + P^- _b &=& P^- _{a'} + P^- _{b'} \label{2} \\
0 & = & \vec p _{a' \perp} + \vec p _{b' \perp}, \nonumber
\end{eqnarray}
where $ P^+ = E + p_z ,~~~ P^- = E -p_z.$

At high energies:
\begin{equation}
P^- _{a'} \simeq m^2 _{a'} / 2 \mid p_{a' z} \mid ,~~~
P^+ _{b'} \simeq m^2 _{b'} / 2 \mid p_{b' z} \mid .
\end{equation}

Thus, the $P^-_{a'}$ and $P^+_{b'}$ distribution has the form:
\begin{eqnarray}
dW & \sim & dP^- _{a'}
/P^- _{a'} \simeq d m^2 _{a'} /m^2 _{a'} , \nonumber \\
dW & \sim & dP^+ _{b'} /P^+ _{b'} \simeq d m^2 _{b'} /m^2 _{b'} .
\end{eqnarray}
The limits of $P^-_{a'}$ and $P^+_{b'}$ are defined as
\begin{equation}
[ P^- _{a}, P^- _{b}],~~~ [ P^+ _{b}, P^+ _{a}]. \label{3}
\end{equation}

In case of hadron-nucleus interactions the kinematics governed by Eqs. (\ref{1}) -- (\ref{3})
is applied for the first collision of the projectile hadron with one of the target nucleons
($a + N_1 \rightarrow a' + N'_1$). For the second collision
($a' + N_2 \rightarrow a'' +N'_2$), analogous relations are used but (\ref{3}) is
changed by
\begin{equation}
[ P^- _{a'}, P^-_{N_2}],~~~ [ P^+ _{N_2}, P^+ _{a'}].
\end{equation}
As a result, the sequence of the collisions leads to a systematic increasing
of the mass of the hadron $a$ if transferred momenta are small.

A similar approach is also applied to simulate nucleus-nucleus interactions. Here
the reactions $a' + b' \rightarrow a''+ b''$ are considered. The distributions
on $P^-_{a''}$ and $P^+_{b''}$ are the same as those for $P^-_{a'}$ and $P^+_{b'}$,
but the limits of $P^-_{a''}$ and $P^+_{b''}$ are redefined as
\begin{equation}
[ P^- _{a'}, P^- _{b'}],~~~ [ P^+ _{b'}, P^+ _{a'}].
\end{equation}

Thus, the model considers the interactions of the cascade particles between themselves
from the point of the view of the cascade-evaporation model. Probabilities of the
multiple interactions are calculated in the Glauber approach.

The excited hadrons are treated as QCD-strings, and a corresponding algorithm is applied
for a simulation of their fragmentation into observed hadrons. Due to the increasing of the
masses, the multiplicities are increased too. Thus, the two factors -- multiple interactions and
the increasing of the masses, explain the increase of the produced particle multiplicity at
a passage from hadron-nucleon interactions to the hadron-nucleus and nucleus-nucleus ones.

In the version of the model code installed at the HepWeb server we take into account the elastic
scattering of hadrons as well as the inelastic interactions (for details see Ref. \cite{Pak}).

The cascading of the produced hadrons in nuclei was not considered in the model. As a results,
a multiplicity of slow particles associated with a destruction of residual nuclei was not
described sufficiently well. In order to overcome the drawback it was proposed in Refs.
\cite{UzFri, Adamovich} to enlarge the Frition model by the reggeon theory inspired model
of the nuclear destruction (RTIM) \cite{Khaled1}.

A simulation of the destruction consists from two stages. At the first stage a set
of interacting nucleons (wounded nucleons) is determined with a help of the Glauber
approximation. At the second stage non-interacting nuclear nucleons are considered.
It is assumed that a non-interacting nucleon located at a relative impact parameter
distance, $r$, from a wounded nucleon can be involved in the interactions with a
probability
$$
W = C_{nd} e^{-r^2/r_{nd}^2},
$$
where $C_{nd}$ and $r_{nd}$ are parameters. The involved nucleon can involve another
spectator nucleon, and so on. It is assumed also that all wounded and involved nucleons
leave the nucleus. Good results have been obtained for light nuclei at $C_{nd}=1$ and
$r_{nd}=1.2$ fm. We recommend to use $C_{nd}=0.2$ and $r_{nd}=1$ fm for heavy nuclei.

To ascribe momenta of the escaped nucleon, we use the algorithm proposed in Ref. \cite{Adamovich}.
To explain its feature let us consider a reaction of a compound system, $(1,2)$, with
a hadron $h$: $(1,2) + h \rightarrow 1 + 2 +h$. Neglecting transverse momenta, a final
state of the reaction will be fully determined by a value of merely one independent
kinematical variable. As the variable let us take
$$
x^+ _1 = (E_1 + p_1)/(E_1 + E_2 + p_1 + p_2).
$$
It is obvious that an analogous variable for the second particle, $x^+_2$, will satisfy
the condition -- $x^+_1+x^+_2=1$. All the other kinematical variables can be determined
from the energy-momentum conservation law.

In a case of a dissociation of two compound systems, $A$ and $B$, containing $A$ and $B$
constituents, respectively, the $i$-th constituent of system $A$ will be described by
$$
x^+ _i =(E_{Ai} + p_{iz})/W^+ _A~~~ and~~~\vec p _{i\perp},
$$
and the $j$-th constituent of system $B$ -- by
$$
y^- _j =(E_{Bj} - q_{jz})/W^- _B~~~ and~~~\vec q _{i\perp}.
$$
Here, $E_{A_i} (E_{B_i})$ and $\vec p_i (\vec q_i )$ are energy and transverse momentum
of $i$-th constituent of $A(B)$.
$$
W^+ _A = \sum ^A _{i=1} (E_{Ai} + p_{iz}),~~~
W^- _B = \sum ^B _{i=1} (E_{Bi} - q_{iz}).
$$

Using the variables, let us write the conservation law as
\begin{eqnarray}
\frac{W^+ _A}{2} + \frac {1}{2 W^+ _A} \sum ^A _{i=1} \frac {m^2_{i\perp}}{x^+ _i} +
\frac{W^- _B}{2} + \frac {1}{2 W^- _B} \sum ^B _{i=1} \frac {\mu ^2_{i\perp}}{y^- _i}
= E^0 _A + E^0 _B,     \nonumber \\
\frac{W^+ _A}{2} - \frac {1}{2 W^+ _A} \sum ^A _{i=1} \frac {m^2_{i\perp}}{x^+ _i} -
\frac{W^- _B}{2} + \frac {1}{2 W^- _B} \sum ^B _{i=1} \frac {\mu ^2_{i\perp}}{y^- _i}
= P^0 _A + P^0 _B, \label{SYS} \\
\sum ^A _{i=1} \vec p _{i\perp} + \sum ^B _{i=1} \vec q _{i\perp} =0,
\nonumber
\end{eqnarray}
where
$m^2 _{i\perp}=m^2 _i + \vec p _{i\perp} ^2,\ \mu ^2 _{i\perp}=\mu^2 _i + \vec q _{i\perp} ^2$,
and $m_i (\mu _i )$ is  a mass of the $i-$th constituent of system $A(B)$.

The system (\ref{SYS}) allows one to determine $W^+ _A, W^- _B$ and kinematic properties of all the particles
in the finite sets
$\{x^+ _i , \vec p _{i\perp} \}$, $\{ y^- _i , \vec q _{i\perp} \}$.
\begin{eqnarray}
W^+ _A =(W^- _0 W^+ _0 + \alpha - \beta + \sqrt{\Delta})/2W^- _0; \\
W^- _B =(W^- _0 W^+ _0 - \alpha + \beta + \sqrt{\Delta})/2W^+ _0; \\
W^+ _0=(E^0 _A + E^0 _B) + (P^0 _{Az} + P^0 _{Bz}); \nonumber \\
W^- _0=(E^0 _A + E^0 _B) - (P^0 _{Az} + P^0 _{Bz}); \nonumber \\
\alpha =\sum ^A _{i=1} \frac {m^2 _{i\perp}}{x^+ _i},~~~
\beta =\sum ^B _{i=1} \frac {\mu ^2 _{i\perp}}{y^- _i}; \nonumber \\
\Delta =(W^- _0 W^+ _0 )^2 + \alpha ^2 + \beta ^2 - 2 W^- _0 W^+ _0
\alpha - 2 W^- _0 W^+ _0 \beta - 2 \alpha \beta ; \nonumber
\end{eqnarray}
$$
p_{iz}=(W^+ _A x^+ _i - \frac {m^2 _{i\perp}}{x^+ _i W^+ _A})/2;~~~
q_{iz}=-(W^- _B y^- _i - \frac {\mu ^2 _{i\perp}}{y^- _i W^- _B})/2.
$$

According to experimental observations \cite{PTfrag}, the average transverse momentum
of a spectator fragment with mass number F obeys to the parabolic law:
$$
<P_{\perp} ^2 >=\frac {F (A-F)}{A} <p_{\perp} ^2 >,~~~ \sqrt{<p_{\perp}^2 >} = 0.05~~ GeV/c.
$$
To reproduce the result, the values of $\vec p _{i\perp}$ for the knocked-out nucleons were
sampled according to the distribution:
\begin{equation}
dW \propto exp(- \vec p _{i\perp} ^2 /
<p_{\perp} ^2 >) d^2 p_{i\perp}, \sqrt{<p_{\perp} ^2 >}=0.05.
\label{pti}
\end{equation}
The sum of the transverse momenta (with "minus" sign) was ascribed to the residual nucleus.

The choose of $x^+_i$ is carried out by
\begin{equation}
dW \propto exp[- (x^+ _i -1/A)^2/(d_x /A)^2 ] d x^+ _i ,~~~d_x =0.05.
\label{xplus}
\end{equation}
The dispersion of the distribution was defined by fitting the average
emission angle of $b$-particles. $x^+$ of the residual nucleus was included
as $1 - \sum  x^+ _i $ .

The knocked-out nucleons occurring in the zone of active action of nuclei
was assumed to change again their characteristics. The new values of
$x^+_i$ and $p_{i\perp}$ were simulated using the distributions (\ref{pti})
and (\ref{xplus}) at
$<p_{\perp} ^2 > = 0.3$ $(GeV/c)^2$ and $d_x=0.21$.
The results from \cite{Propan1, Propan2, ABaldin} were used for
a determination of the fitting parameters.

The calculation of the nuclear residual excitation energy is carried out according to
Ref. \cite{Abul-Magd}. The implemented method is described in details in Ref. \cite{Adamovich}.
For a simulation of a relaxation of the excited nuclei the standard evaporation model
is used \cite{Weis37}.

A version of the model applied in Refs. \cite{Adamovich,AC} underestimated negative
charged particle multiplicities especially in the target fragmentation region. Thus,
an attempt to take into account non-nucleon degree of freedom in nuclei has been considered.

It is no doubt that the nucleons in nuclei can have virtual transitions --
$N \rightarrow N+\pi$ and $N+\pi \rightarrow N$. The virtual pairs can
fall onto the mass-shell and become real ones due to an interaction with a
projectile hadron. A calculation of cross sections and properties of the
processes requires a solution of numerous questions of the nuclear theory,
and can not be done in a full volume in the present time. Additionally
it is needed to consider a creation of the $N\pi$ pairs in the reggeon cascade.
Thus, for a qualitative understanding of processes we have assumed that an
involved or wounded nucleon can be the $N\pi$ pair with a probability of $\sim 20$\%.
More concrete, we assume that a proton or a neutron participated in the interaction
can be $\Delta ^+$- or $\Delta ^0$ isobar. It is the most simple solution from
a programming point of view. It throws away a complicated question on kinematical
properties of the pairs.

\section*{Appendix: HIJING and AMPT Models}
The quark-gluon degree of freedom manifests itself directly at super high energies in
hadron-hadron interactions -- jets of hadrons are produced which are interpreted as
results of fragmentations of quarks and gluons suffered hard collisions with large
transverse momenta. Inclusive properties of the jets are described by
the quantum chromo-dynamics (QCD) with a high precision. The most popular Monte Carlo codes
for a simalation of the processes are PYTHIA \cite{Pythia} and HERWIG \cite{Herwig}.
Because the hadronisation process - the fragmentation of quarks and gluons into hadrons,
can not be described in the frame of QCD, various phenomenological models are applied in
the case. The total and elastic hadron-hadron cross sections are not described by QCD also.
They are determined by the so-called soft interactions with low transverse momenta. Thus,
for a complete interpretation of experimental data various models of a coupling of the
hard and soft processes were proposed. One of the successful model is the HIJING code
(Heavy Ion Jet Interaction Generator \cite{Hijing1,Hijing2}).

The model is a generalization of the Fritiof model. It considers the hard jet production
simulated with a help of the Pythia code. It is assumed in the model that the inelastic
nucleon-nucleon interaction cross section has a form:
\begin{equation}
\sigma^{in}=\int d^2b~\left[1-e^{-(\sigma_{soft}+\sigma_{hard})T_N(\vec b)}\right]=
\label{Sin}
\end{equation}
$$
=\int d^2b~\left( \left[1-e^{-\sigma_{soft}~T_N(\vec b)}\right]~e^{-\sigma_{hard}~T_N(\vec b)}
~+~\sum_{j=1}\frac{[\sigma_{hard}~T_N(\vec b)]^j}{j!}
e^{-\sigma_{hard}~T_N(\vec b)}\right),
$$
where $T_N$ is an overlap function of partons (quarks and gluons) from colliding hadrons
at impact parameter  $\vec b$. It is given by the Fourier transform of the dipole form
factor of the proton.
$$
T_N(\vec b,s)=2\frac{\chi_0(\xi)}{\sigma_{soft}(s)},~
\chi_0(\xi)=\frac{\mu_0^2}{96}(\mu_0\xi)^3 K_3(\mu_0\xi),~
\xi=|\vec b|/b_0(s),
$$
$$
\mu=3.9,~~~\pi\ b^2_0=\sigma_{soft}(s)/2,~~~\sigma_{soft}(s)=57~(mb).
$$
$K_3$ -- Mac-Donald function, $\sigma_{hard}$ -- integrated inclusive jet production
cross section given by QCD.

Eq. \ref{Sin} allows one to determine a number of the hard scatterings accompanied
all time by the excitation of hadron residues. If the number is equal to zero, the soft
interaction is simulated as in the Fritiof model.
For each hard interaction the kinetic variables of the
scattered partons (gluons) are determined by calling PYTHIA \cite{Pythia} subroutines.
Since jet production is dominated by gluon scatterings, it is assumed that
quark scatterings only involve valence quarks, and hard processes are restricted
only by the gluon-gluon scatterings. Simplification is also made
for the color flow in the case of multiple jet production. Produced gluons are
ordered in their rapidities and then connected with their parent valence quarks
or diquarks to form string systems. Finally, fragmentation subroutine of JETSET
is called for hadronization.

Eq. \ref{Sin} is entering the Glauber formulae for hadron-nucleus and
nucleus-nucleus cross sections in the corresponding routines of the HIJING code.
Besides a change of the gluon distribution in nuclei (the nuclear shadowing)
and interactions of the scattered gluons with spectators nucleons
(the final state interactions, or jet quenching) are taken into account.

According to the Glauber approximation, it is assumed that a nucleus-nucleus
collision can be decomposed into binary nucleon-nucleon collisions which generally
involve the wounded nucleons. In a string picture, the wounded nucleons become
strings excited along the beam direction. At high energy, the excited strings are
assumed to interact again like the ordinary nucleon-nucleon collisions before they
fragment. Unlike the Fritiof model, it is allowed that an excited string can be
de-excited within the kinematic limits in the subsequent collisions. The binary
approximation can also be applied to rare hard scatterings which involve only
independent pairs of partons. The probability for a given parton to suffer multiple
high $P_T$ scatterings is small and is not implemented in the current version of the
program. A three-parameter Wood-Saxon nuclear density is used for a calculation of
the number of binary collisions at a given impact parameter. After all binary
collisions, the scattered partons are connected with the corresponding valence
quarks and diquarks to form string systems. The strings are then fragmented into
particles.

One of the most important nuclear effects in relativistic heavy ion collisions is
the nuclear modification of parton structure functions. It has been observed \cite{EMC}
that the effective number of quarks and antiquarks in a nucleus is depleted in the
low region of x. This is called in the code as the nuclear shadowing, and is implemented.

Another important nuclear effect on the jet production in heavy ion collisions is
the final state integration. In high-energy heavy ion collisions, a dense hadronic
or partonic matter must be produced in the central region. Because this matter can
extend over a transverse dimension of at least $R_A$, jets with large $P_T$ from hard
scatterings have to traverse this hot environment. Thus it is very important to
simulate the interaction of jets with the matter and the energy loss they suffer.
It is estimated \cite{JetQ} that the gluon bremsstrahlung induced by soft interaction
dominate the energy loss mechanism. The induced radiation in HIGING is modeled via
a simple collinear gluon splitting scheme with given energy loss $dE/dz$. It is
assumed also that the interaction only occur with the locally comoving matter in
the transverse direction. The interaction points are determined via a probability
$$
dP = \frac{dl}{\lambda_s}e^{-l/\lambda_s},
$$
with given mean free path $\lambda_s$, where $l$ is the distance the jet has
traveled after its last interaction. The induced radiation is simulated by
transferring a part of the jet energy $\Delta E(l) = ldE/z$ as a gluon kink to
the other string which the jet interacts with. The procedure is continued until
the jet is out of the whole excited system or when the jet energy is smaller than
a cutoff below which a jet can not loss energy any more. The cutoff is the same
as the cutoff $P_0$ for jet production. To determine how many and which excited
strings could interact with the jet, it is assumed that excited strings within
a cylinder of radius $r_s$ along the jet direction could interact with the jet.
$\lambda_s$ and $r_s$ are two parameters in of the jet quenching effect.

The code of the HIJING model operates quite fast. Thus it was widely used by
RHIC and LHC experimental collaborations at designs of experimental setups.

The AMPT model (A Multi-Phase Transport model) \cite{AMPT} is built on the HIJING
code. It considers additionally to the HIJING subjects the final state interactions
between the partons. The interactions of the produced hadrons with nuclear matter
is taken into account too.

\section*{Appendix: Calculation of nucleus-nucleus interaction cross sections at
                    high energy in the Glauber approach}
The amplitude of scattering a nucleus $A$ on a nucleus $B$, when each of
them transforms from the initial state $|i\rangle$ into the final states
$|f\rangle$, is given in the Glauber's approach \cite{Franco},\cite{Czyz}--
\cite{Formanek} by the expression
\begin{equation}
F_{A,B}(\vec q) = \frac{i~ p_A}{2\pi}
\int d^2b ~ e^{\imath \vec q ~ \vec b}
\langle{f_A,f_B}| 1-\prod_{j=1}^A \prod_{k=1}^B
(1-\gamma(\vec b-{\vec s}_j+{\vec {\tau}}_k))
|i_B,i_A\rangle ,
\label{eq3}
\end{equation}
\noindent where $p_{A}$ is the momentum of the projectile nucleus $A$,~
$\vec q$ is the transferred transverse momentum, $\vec b$ is the impact
parameter, ~$\gamma$ is an amplitude of elastic $NN$ scattering in the
impact parameter representation,~$\lbrace{\vec s_j\rbrace},~j~=~$1, 2, ...,A
and ~$\{\vec {\tau}_k\},~k~=1,2,...,B$~ are coordinates of nucleons
within, respectively, $A$ and $B$ nuclei on the impact parameter
plane. These coordinates are measured from the center of mass of each
nucleus respectively, too.

Starting from eq. \ref{eq3}, it is possible to find $AB$ elastic scattering
amplitude
\begin{eqnarray}
F_{AB}^{el}(\vec q)=
\frac{i~  p_A}{2\pi}\int d^{2}b ~
e^{\imath \vec q ~ \vec b} ~
\{1-\prod_{j=1}^{A}\prod_{k=1}^{B}
(1-\gamma (\vec b-{\vec s}_j+{\vec {\tau}}_k))\}
\cdot   \label{eq4} \\
\qquad
{\left|{\psi}_A({\vec r}_1,\ldots,{\vec r}_A)\right|}^2
~
{\left|{\psi}_B({\vec t}_1,\ldots,{\vec t}_B)\right|}^2
\prod_{j=1}^{A}d^{3}r_{j}~\prod_{k=1}^{B}d^{3}t_k , \nonumber
\end{eqnarray}

\noindent differential cross section

\begin{equation}
\frac{d \sigma^{el}}{d\Omega}={\Bigl|F_{AB}^{el}\Bigr|}^2, \label{eq5}
\end{equation}

\noindent and total cross section

\begin{equation}
\sigma_{AB}^{tot}=\frac{4\pi}{p_A}Im F_{AB}^{el}(0) . \label{eq6}
\end{equation}

The cross section of the quasi-elastic scattering of the nucleus $A$ when it
is conserved, but other nucleus $B$ undergoes all excitations
including destruction too $( A + B\rightarrow A + X )$ is given by
the expression
\begin{eqnarray}
\sigma (A+B \to A+X) =
\int d^2 b ~
\{
1-\prod_{i=1}^A\prod_{k=1}^B ~
(1-\gamma (\vec b-{\vec s}_i+{\vec \tau}_k)) ~
\}
\cdot            \label{eq7}\\ \nonumber
\{
1-\prod_{j=1}^A \prod_{k=1}^B ~
(1-\gamma ^*(\vec b-{\vec s}_i'+{\vec \tau}_j))
\}
\cdot \\ \nonumber
{\left|{\psi}_A({\vec r}_1,\ldots,{\vec r}_A)\right|}^{2}
{\left|{\psi}_A({\vec r}_{1}',\ldots {\vec r}_{A}')\right|}^{2}
{\left|{\psi}_B({\vec t}_1,\ldots,{\vec t}_B)\right|}^{2}
\cdot \\ \nonumber
\prod_{i=1}^{A}d^3 r_{i} \prod_{i=1}^{A}d^3 r_i'
\prod_{i=1}^{B}d^3 t_{i}
~-~ \sigma_{AB}^{el} .
\end{eqnarray}

Finally, the cross section of the production of new particles can be
defined as
\begin{eqnarray}
\sigma_{AB}^{prod}=
\int d^2 b ~
\{
1 - \prod_{i=1}^A \prod_{k=1}^B ~
(1-p(\vec b-{\vec s}_i+{\vec \tau}_k))
\}
\cdot \label{eq8} \\
\left|{\psi}_A({\vec r}_1,\ldots,{\vec r}_A)\right|^2
~
\left|{\psi}_B({\vec t}_1,\ldots,{\vec t}_B)\right|^2
\prod_{i=1}^{A}d^3 r_i ~\prod_{i=1}^{B}d^3 t_i    , \nonumber
\end{eqnarray}
$$
p(\vec b)=\gamma(\vec b)+\gamma^*(\vec b)-
\gamma(\vec b) \gamma^{*}(\vec b).
$$

Eq.\ref{eq8} can be re-written in the form where
each of terms would be interpreted as a probability of some process
\begin{eqnarray}
\sigma_{AB}^{prod}=
\int d^2 b ~
\left\{
\sum_{i=1}^A \sum_{j=1}^B ~
\frac{p({\vec b}-{\vec s}_i+{\vec \tau}_j)}
     {1-p({\vec b}-{\vec s}_i+{\vec \tau}_j)}
\prod_{k=1}^A \prod_{l=1}^B ~
(1-p({\vec b}-{\vec s}_k+{\vec \tau}_l))
\right.
  \label{eq9} \\
+\frac{1}{2} \cdot
\sum_{{i=1,j=1}\atop {\imath \ne \jmath}}^A \sum_{k=1}^B ~
\frac{p({\vec b}-{\vec s}_i+{\vec \tau}_k)}
{1-p({\vec b}-{\vec s}_i+{\vec \tau}_k)}              
\frac{p({\vec b}-{\vec s}_j+{\vec \tau}_k)}
{1-p({\vec b}-{\vec s}_j+{\vec \tau}_k)}   ~
\nonumber \\
\cdot \left.
\prod_{l=1}^A \prod_{m=1}^B ~
(1-p({\vec b}-{\vec s}_l+{\vec \tau}_m)) +{\ldots}
\right\}
 \nonumber \\
\cdot
\left|{\psi}_A({\vec r}_1,\ldots,{\vec r}_A)\right|^2
\left|{\psi}_B({\vec t}_1,\ldots,{\vec t}_B)\right|^2
\prod_{i=1}^{A}d^3 r_i ~ \prod_{i=1}^{B}d^3 t_i .   \nonumber
\end{eqnarray}

\noindent Here the first term in the first braces is interpreted as a
probability that the only one inelastic collision between $i$-th
nucleon from nucleus $A$ and $j$-th nucleon from nucleus $B$
takes place when all nucleons coordinates are fixed. The second term
describes a probability of inelastic collision of the $k$-th nucleon
from nucleus $B$ with $i$-th and $j$-th nucleons in $A$ nucleus,
etc. Nucleons involved in the collisions are called as "wounded" ones, but
the others are called as "spectators" ones.

To calculate all cross sections discussed above, it is necessary
to have a function $\gamma(\vec b)$ and the square of the modulus of the ground state
wave function $\left|\psi\right|^2$ of $A$ and $B$ nuclei.

The approximation
\begin{equation}
\gamma (\vec b)= \sigma_{NN}^{tot}\cdot \frac{1-i\alpha}{4\pi~B}
~e^{-\vec b^2/ 2B} \label{eq10}
\end{equation}
\noindent is often used at $E>1$ $GeV/nucleon$. Here $ \sigma_{NN}$ is
the total cross section of $NN$ interaction, $\alpha$ is the ratio of the real part
to the imaginary part of the elastic scattering amplitude at zero momentum
transfer, $B$ is the slope parameter of the differential cross section of
the elastic $NN$ scattering. Expression \ref{eq10} corresponds to the following
parameterization of the elastic $NN$ scattering amplitude in the momentum
representation
$$
f_{NN}(\vec q)=\frac{ip_A}{4\pi }\cdot \sigma_{NN}^{tot}
(1-i\alpha)~e^{-B{\vec q^2}/2} .
$$
\noindent
Sets of values $~\sigma_{NN}^{tot}$, $\alpha$ and $B$ at various
energies are presented  in the compilations \cite{compil}. The Particle Data
Group proposed the following parametrizations for the total and elastic
$NN$ cross--sections at high energies \cite{PDG}:
\begin{equation}
\sigma (p)= A ~+~ Bp^n ~+~ C\ln ^2(p) ~+~ D\ln (p), \label{eq11}
\end{equation}
\noindent
where $p$ is the laboratory momentum in $GeV/c$. The parameters -- $A$, $B$,
$C$, and $D$ are given in Ref. \cite{PDG}.

Function $\left|{\psi}_A\right|^2$ is often given as
\begin{equation}
\left|{\psi}_A\right|^2 = \prod_{i=1}^{A}\rho_{A}({\vec r}_i) , \label{eq12}
\end{equation}
\noindent
where $~\rho~$ represents the one-particle density of a nucleus. In this case
the aggregate of the nucleon coordinates does not meet self-evident demands
\begin{equation}
\sum_{i=1}^{A}{\vec r}_i=0 . \label{eq13}
\end{equation}
\noindent
Taking into account this condition is named "an account of
center of mass correlation".

If projectile is a nucleon, it is
necessary to set a density $\rho_A$ as a delta-function
$~\rho_{A}(\vec r)=\delta(\vec r)$ in this case.

The parameterization from Ref. \cite{Azhgir} is used for deuteron
\begin{equation}
\left|{\psi }_d\right|^2 = \sum_{i=1}^3 c_{i}
e^{-\vec r^2/4 \gamma_i} , \label{eq14}
\end{equation}

\begin{eqnarray}
\gamma_{1}=225~(GeV/c)^{-2},~~~c_1=0.178/(4\pi \gamma_1)^{3/2},\nonumber\\
\gamma_{2}=~45~(GeV/c)^{-2},~~~c_1=0.287/(4\pi \gamma_2)^{3/2},\nonumber\\
\gamma_{3}=~25~(GeV/c)^{-2},~~~c_1=0.535/(4\pi \gamma_3)^{3/2}.\nonumber
\end{eqnarray}
\noindent
Here $r$ is a distance between a proton and a neutron within a deuteron.

For ${}^3$H,~${}^3$He~and~${}^4$He~nuclei $|\psi_A|^2$ has been chosen as
\begin{equation}
|\psi_A|^2 = \delta (\frac{1}{A}\sum_{\imath=1}^{A}{\vec r}_i)
\prod_{i=1}^A
\frac{1}{{(\pi R_A^2)}^{3/2}}e^{-{r_i^2}/{R_A^2}}, \label{eq15}
\end{equation}
\noindent
where
$R_{{}^{3}H} = R_{{}^{3}He} = 1.81 ~ fm,~
R_{{}^{4}He} = 1.37 ~ fm$.
For all other nuclei ($A\ge 6$) a one-particle density has been defined
as
\begin{equation}
\rho_{A}(r)=~const~/~{(1+e^\frac{r-R_A}{c})}, \label{eq16}
\end{equation}
\noindent with $R_{A} = 1.07\cdot A^{1/3} ~ fm,~ c = 0.545 ~ fm$.
The center of mass correlation has been taken as in paper \cite{CMCorr}.

The calculation of the cross sections is performed in accordance with the
method discussed in \cite{DIAGEN}, where the statement
$$
|\psi_A|^{2}~|\psi_B|^{2}
\prod_{i=1}^{A}d^3 r_i \prod_{j=1}^{B}d^3 t_j
$$
is treated as a probability measure to find out different sets of
nucleon coordinates in $A$ or $B$ nucleus. In this case, the
cross sections are found as mean values over various sets of
nucleon coordinates.  Thus it can be written in the form
\begin{equation}
\sigma_{AB}^{tot} =
\frac{1}{N_{stat}}\cdot\sum_{i=1}^{N_{stat}} \int
d^2 b ~ E_{AB}^{tot}(\vec b,\{\vec {r_A}\},\{\vec {t_B}\}) \label{eq17}
\end{equation}
where
$$
E_{AB}^{tot} = 1-\prod_{i=1}^{A}\prod_{j=1}^{B}
(1-\gamma( \vec b-\vec {s_i}+\vec {\tau_j} )) ,
$$
and $N_{stat}$ denotes a number of various sets of nucleon
coordinates. Therefore, the accuracy of the calculation depends on the
value of $N_{stat}$. (The expressions analogous to \ref{eq17} can be written for all
other cross sections.)

The function $E$ is called as a distribution over impact
parameter for inelastic $AB$ interactions:
\begin{equation}
P(\vec b) =
\frac{1}{N_{stat}} \cdot
\sum_{i=1}^{N_{stat}}
E_{AB}^{prod} (\vec b,\{\vec {r_A}\},\{\vec {t_B}\}), \label{eq18}
\end{equation}
where
$$
E_{AB}^{prod} = 1-\prod_{i=1}^{A}\prod_{j=1}^{B}
(1-p( \vec b-\vec {s_i}+\vec {\tau_j} )).
$$

In accordance with the expression \ref{eq9}, the algorithm for simulation of
inelastic collisions consists of the following steps:
\begin{enumerate}
\item Calculation and tabulation of the function $P(\vec b)$;

\item Generation of the impact parameter ${\vec b}$ in accordance with
$P(\vec b)$ distribution;

\item Sampling of the coordinates of the nucleons within nuclei
according to $|\psi_{A}|^{2}$ and  $|\psi_{B}|^{2}$ distribution
functions;

\item Sampling and storing pairs of nucleons interacting inelastically.
For this purposes $A\cdot B$ random
numbers $\xi_{ij}~( i=1,\ldots,A,~j=1,\ldots,B )$, uniformly
distributed in the [0,1.] interval, are chosen. If
$\xi_{ij}<p(\vec b-{\vec s}_i+\vec {\tau_j})$
then it is considered that an inelastic
collision takes place between $i$-th nucleon of $A$ nucleus
and $j$-th nucleon of $B$ nucleus.

\end{enumerate}

The first step is performed only once with given nuclei $A$ and $B$.
The steps 2--4 are repeated as many times as it is necessary.

The method described above has been realized as a set of FORTRAN routines.
These routines and their assignments are presented in the table.

\noindent$*****************************************************$

\begin{tabular}{llclc}
NN & {\bf name} &*& {\bf Routines'~function} & {\bf Subject} \\ \hline
1  & TOTAL      &*& Calc. total cross sect. & $\sigma_{AB}^{tot}$\\
2  & ELASTIC    &*& Calc. elastic cross sect.& $\sigma_{el}$      \\
3  & AB\_TO\_AX &*& Calc. cross sect. of reaction & $A+B\to A+X$ \\
4  & AB\_TO\_XB &*& Calc. cross sect. of reaction & $A+B\to X+B$ \\
5  & AB\_TO\_X  &*& Calc. cross sect. of reaction & $A+B\to X $  \\
   &            &*& without process of produc-    &              \\
   &            &*& tion of particles             &              \\
6  & PRODUCT    &*& Calc. cross sect. of multi--~ &$\sigma_{AB}^{prod}$\\
   &            &*& particle production            &              \\
7  & G\_X\_SECT &*& Calc. all above cross sect.   &              \\
8  & GL\_STAR   &*& Simulates    inelastic events &              \\
\end{tabular}

\noindent$*****************************************************$

All of them can be called through the HepWeb server.


\begin{thebibliography}{1111}
\bibitem{NRV}V. Zagrebaev and A.Kozhin, "Nuclear Reactions Video (knowledge
base on low energy nuclear physics)", JINR Report No. E10-99-151, Dubna, 1999;
http://nrv.jinr.ru

\bibitem{JetWeb}J.M. Butterworth and S. Butterworth, arXiv: hep-ph/0210404 (2002).

\bibitem{Pythia}http:/{/}home.thep.lu.se/$\sim$torbjorn/Pythia.html

\bibitem{Herwig}http:/{/}hepwww.rl.ac.uk/theory/seymour/herwig/

\bibitem{HZTOOL}J. Bromley et al, Hamburg 1995/96, Future physics at HERA, vol. 1, 611-612.

\bibitem{JAS}A.S. Johnson, CHEP 2000, Computing in high-energy and nuclear physics,
741-745. http://jas.freehep.org/

\bibitem{DTD}http://www-sldnt.slac.stanford.edu/jas/Documentation/
howto/xml/default.shtml

\bibitem{JDBC}M. Matthews, http://mmmysql.sourceforge.net/old-index.html;\\
http://www.mysql.com/downloads/api-jdbc.html

\bibitem{Tomcat}Tomcat http://jakarta.apache.org/tomcat

\bibitem{QTsummed}J.C. Collins, D.E. Soper and G. Sterman, Nucl. Phys. {\bf B250}
(1985) 199;\\
F. Landry, R. Brock, Pavel M. Nadolsky, C.P. Yuan, Phys. Rev. {\bf D67} (2003) 073016,
e-Print: hep-ph/0212159.

\bibitem{UrQMD1}S.A. Bass et al., Prog. Part. Nucl. Phys. {\bf 41} (1998) 225;
nucl-th/9803035.

\bibitem{UrQMD2}M. Bleicher et al., J. Phys. {\bf G25} (1999) 1859; hep-ph/9909407.

\bibitem{CMB-home-page}http:/{/}www.gsi.de/fair/experiments/CBM/index\_e.html

\bibitem{PANDA-home-page}http:/{/}www-panda.gsi.de/auto/\_home.htm

\bibitem{RQMD}H. Sorge, Phys. Rev. {\bf C52} (1995) 3291.

\bibitem{Hijing1}X.-N. Wang and M. Gyulassy, Phys. Rev. {\bf D44} (1991) 3501.

\bibitem{Hijing2}M. Gyulassy and X.-N. Wang, Comput. Phys. Commun. {\bf 83} (1994) 307;
                nucl-th/9502021.

\bibitem{GEANT}http:/{/}geant4.web.cern.ch/geant4/

\bibitem{CBM_TPR}CBM Collaboration, {\it  Compressed Baryonic Matter Experiment -
Technical Status Report}, 2005.

\bibitem{PANDA_TPR}PANDA Collaboration, {\it Technical Progress Report for:
PANDA (AntiProton Annihilations at Darmstadt)
Strong Interaction Studies with Antiprotons}, 2005.

\bibitem{Zhenis}V.S.~Barashenkov, F.G.~Zheregy and Zh.Zh.~Musulmanbekov, Preprint JINR
{\bf P2-83-117} (1983) Dubna.

\bibitem{Fritiof}B. Andersson et al., Nucl. Phys. {\bf B281} (1987) 289;
B. Nilsson-Almquist and E. Stenlund,  Comp. Phys. Commun. {\bf 43} (1987) 387.

\bibitem{cascade1}V.S.~Barashenkov and V.D.~Toneev, "Interaction of high energy
particles and atomic nuclei with nuclei", Moscow, Atomizadt, 1972.
\bibitem{cascade2}N.W. Bertini et al., Phys. Rev. {\bf C9} (1974) 522.
\bibitem{cascade3}N.W.~Bertini et al., Phys. Rev. {\bf C14} (1976) 590.
\bibitem{cascade4}J.P.~Bondorf et al., Phys. Lett. {\bf 65B} (1976) 217.
\bibitem{cascade5}J.P.~Bondorf et al., Zeit. Phys. {\bf A279} (1976) 385.
\bibitem{cascade6}V.D.~Toneev and K.K.~Gudima, Nucl. Phys. {\bf A400} (1983) 173.


\bibitem{Adamovich}M.I.Adamovich et al. (EMU-01/12 - collab.),
Zeit. Phys. {\bf A 358} (1997) 337.


\bibitem{UzFri}V.V. Uzhinskii, JINR Commun, E2-96-192, Dubna, 1996.

\bibitem{Khaled1}Kh. El-Waged and V.V. Uzhinskii,Phys. Atom. Nucl. {\bf 60} (1997) 828;
 Yad. Fiz. {\bf 60} (1997) 925.

\bibitem{UzhiUrQMD}A.~Galoyan, J.~Ritman and V.~Uzhinsky, nucl-th/0605021 (2006).

\bibitem{Uzhi_Val}A.~Galoyan, J.~Ritman and V.~Uzhinsky, nucl-th/0605039 (2006).


\bibitem{GENSER}http://lcgapp.cern.ch/project/simu/generator/

\bibitem{Uzhi_HIJVal}V.V.~Uzhinsky, hep-ph/0312089 (2003).

\bibitem{compil}O.~Benary, L.R.~Price and G.~Alexander. NN and ND interactions [above 0.5 GeV/c ]
-- a compilation, UCRL-20000NN, 1970;\\
V.~Flaminio, W.G.~Moorhead, D.R.O.~Morrison and N.~Rivoire. Compilation of
cross--sections: p and $\bar p$ induced reactions, CERN--HERA 84-01, 1984.

\bibitem{PDG}The Particle Data Group, Phys. Rev. {\bf D54} (1996) 125.

\bibitem{LITSR2005_1}LIT Scientific Report 2004-2005, 2005-179, JINR, DUBNA 2005, pp. 25-28.

\bibitem{LITSR2005_2}LIT Scientific Report 2004-2005, 2005-179, JINR, DUBNA 2005, pp. 29-32.

\bibitem{Zador}S.Yu.~Shmakov, V.V.~Uzhinskii and A.M.~Zadorozhny,
Comp. Phys. Commun.  {\bf 54} (1989) 125.

\bibitem{DIAGEN}S.Yu. Shmakov, V.V. Uzhinski, A.M. Zadorojny,
Comp. Phys. Commun. 54. (1989) 125.

\bibitem{Galperin}A.G.~Galperin and V.V.~Uzhinskii, JINR preprint {\bf E2-94-505} (1994).

\bibitem{Blann66}M. Blann, Ann. Rev. Nucl. Sci. {\bf 17} (1966) 478.

\bibitem{Weis37}V. Weisskopf, Phys. Rev. {\bf 52} (1937) 295.

\bibitem{Fried83}W.A. Friedman, Phys. Rev. {\bf C28} (1983) 16.

\bibitem{Mashnik94}S.G. Mashnik In "Proceedings of a Specialists
Meeting - Intermediate Energy Nuclear Data: Models and Codes".
Paris, 1994, P.107.

\bibitem{Blann94} Blann M.B., Gruppelaar H., Nagel P., Rodens J.
Report "International Code Comparison for Intermediate Energy Nuclear
Data", NEA, OECD, Paris, 1994.

\bibitem{Bertch}G.F. Bertch and S. Das Gupta, Phys. Rep. {\bf 160} (1988) 189.

\bibitem{Cassing}W. Cassing, V. Metag, V. Mosel, K. Nuta, Phys. Rep. {\bf 188} (1990) 363.

\bibitem{Aichelin}J. Aichelin, Phys. Rep. {\bf 202} (1991) 233.

\bibitem{Sorge1}H. Sorge, H. Stoker, W. Greiner, Ann. of Phys. (N.Y.)
{\bf 192} (1989) 266; Nucl. Phys. {\bf A498} (1989) 567c.

\bibitem{Sorge2}H. Sorge et al., Zeit. fur Phys. {\bf C47} (1990) 629.

\bibitem{Sorge3}H. Sorge et al., Zeit. fur Phys. {\bf C59} (1993) 85.

\bibitem{PDG92} K. Hikasa et al., Phys. Rev. {\bf D45} (1992) III.83.

\bibitem{donna}A. Donnachie and P.V. Landshoff, Phys. Lett. {\bf B296} (1992) 227.

\bibitem{kochP89a}P. Koch, B. Muller and J. Rafelski, Phys. Rept. {\bf 142} (1986) 167.

\bibitem{andersson83a} Bo Andersson, G. Gustafson and B. Soderberg,
Z. Phys. {\bf C20} (1983) 317.

\bibitem{field77a} R.D. Field and R.P. Feynman, Phys. Rev. {D15} (1977) 2590.

\bibitem{field78a} R.D. Field and R.P. Feynman, Nucl. Phys. {\bf B136} (1978) 1.

\bibitem{Gana}
B. Ganhuyag and V. Uzhinsky, JINR Commun. {\bf ³1-97-315}, Dubna, 1997;
JINR Commun. {\bf ³2-97-397}, Dubna, 1997;
B. Ganhuyag, JINR Commun. {\bf ³2-98-26}, Dubna, 1998.

\bibitem{Pak}
V.V. Uzhinsky and A.S. Pak, Yad. Fiz. {\bf 59} (1996) 1109.

\bibitem{PTfrag}A.I. Bondarenko, V.V. Rusakova, J.A. Salomov, G.M. Chernov,
Yad. Fiz. {\bf 55} (1992) 135 (Sov. J. Nucl. Phys. {\bf 55} (1992)).

\bibitem{Propan1} D. Armutlisky et al., Zeit. fur Phys. {\bf A328}
(1987) 455.

\bibitem{Propan2} H.N. Agakishiev et al., Yad. Fiz. {\bf 56} (1993) 170
(Sov. J. Nucl. Phys. {\bf 56} (1993)).

\bibitem{ABaldin}A.A. Baldin, Yad. Fiz. {\bf 56} (1993) 174 (Sov. J.
Nucl. Phys. {\bf 56} (1993)).

\bibitem{Abul-Magd}A.Y. Abul-Magd, W.A. Friedman, J. Hufner
Phys. Rev. C34 (1986) 113.

\bibitem{AC} A.I. Bondarenko et al., Yad. Fiz. {\bf 65} (2002) 95.

\bibitem{EMC}EM Collab., J. Ashman et al., Phys. Lett. {\bf B202} (1988) 603;
EM ColIab., M. Arneodo et al., Phys. Left. {\bf B211} (1988) 493.

\bibitem{JetQ}M. Gyuiassy et al., Nucl. Phys. {\bf A538} (1992) 37c;
M. Gyulassy and X-.N. Wang, Nuci. Phys. {\bf B420} (1993) 583.

\bibitem{AMPT}Z.W. Lin et al, Phys. Rev. {\bf C72} (2005) 064901;
Phys. Rev. {\bf C64} (2001) 011902; B. Zhang et al, Phys. Rev. {\bf C61} (2000) 067901.

\bibitem{Franco}V. Franco, Phys. Rev. {\bf 175} (1968) 1376.

\bibitem{Czyz}W. Czyz, L.C. Maximon, Ann. of Phys. (N.Y.) {\bf 52} (1969) 59.

\bibitem{Kofoed}O. Kofoed--Hansen, Nuov. Cim. {\bf 60A} (1969) 621.

\bibitem{Harrington}D.R. Harrington and A. Pagnamenta, Phys. Rev. {\bf 184} (1969) 1908.

\bibitem{Formanek}J. Formanek, Nucl. Phys. {\bf 1312} (1969) 441.

\bibitem{Azhgir} L.S. Azhgirey et al., JINR preprint {\bf E2-12683}, Dubna, 1979.

\bibitem{CMCorr} V.V. Uzhinskii and S.Yu. Shmakov, S.J. Nucl. Phys. {\bf 57} (1994) 1532.

\end{thebibliography}
\end{document}